\pdfoutput=1 
\documentstyle[12pt,graphicx,natbib]{article}
\def\sun{\hbox{$\odot$}}
\def\apj{ApJ\,}
\def\aap{A\&A\,}
\def\aj{AJ}
\def\araa{ARA\&A}
\def\apjs{ApJS}

\def\nat{Nature\,}

\begin{document}
\title{
Semi-analytical formulas 
for the Hertzsprung-Russell Diagram
}
\author
{L. Zaninetti             \\
Dipartimento di Fisica Generale, \\
           Via Pietro Giuria 1   \\
           10125 Torino, Italy   
}

\maketitle
\section*{}
The absolute visual magnitude as function of the observed
color (B-V) , also named Hertzsprung-Russell diagram
can be described through five equations;
that in presence of calibrated stars means 
eight  constants.
The developed framework allows to deduce the remaining
physical parameters that are  mass , radius and luminosity.
This new technique is applied to  the first 10 PC ,
the first 50 pc  ,
the Hyades       and to the determination 
of the  distance of a cluster.
The case of  the white dwarfs
is  analyzed  assuming  the  absence  of calibrated data:
our  equation produces a smaller $\chi^2$ 
in respect to the standard color-magnitude calibration
when  applied  to the  Villanova Catalog of Spectroscopically 
Identified White Dwarfs.
The theoretical  basis of the 
formulas  for the colors and the bolometric correction
of the stars are clarified  
through   a Taylor expansion in the temperature 
of  the Planck distribution.
\\
keywords                    \\
         {stars: formation       ;
          stars: statistics      ;
          methods: data analysis ;
          techniques: photometric }

\section{Introduction}
The diagrams in absolute visual magnitude versus spectral type for
the stars started
with~\cite{Hertzsprung_1905,Rosenberg_1911,Hertzsprung_1911}~. The
original Russell version can be found 
in \cite{Russell_1914_a,Russell1914Nature,Russell_1914_b},  in the
following H-R diagram. The common  explanation is through the
stellar evolution, see for example chapter~VII
in~\cite{Chandrasekhar_1967}. Actually the presence of
uncertainties in the stellar evolution makes the comparison
between theory and observations an open field of research ,
see~\cite{Maeder1984,Madore1985,Renzini1988,Chiosi1992,Bedding1998}.
Modern application of the H-R diagram 
can  be found    in \cite{Hyades2001} applied to the  Hyades when 
 the parallaxes are provided by  Hipparcos, and 
in ~\cite{Al-Wardat2007} applied 
to the binary systems COU1289 and COU1291.

The Vogt theorem , see~\cite{Vogt_1926},
 states that

{\bf Theorem 1~~}{\it The structure of a star is determined by it's mass and it's
chemical composition.}

Another approach  is the parametrization of physical quantities
such as absolute magnitude, mass , luminosity
and radius as a function of the temperature ,
see for example~\cite{cox} for  Morgan and Keenan
classification , in the following MK, see~\cite{MK1973}.
The temperature is not an observable quantity  and therefore
the parametrization of the observable and not observable quantities
of the stars as a function of the observable color is an
open problem in astronomy.

{\bf Conjecture 1~~} {\it The absolute visual magnitude $M_V$    is a
function, $F$,  of the selected color
\begin{eqnarray}
M_V= F( c_1, ....., c_8 ,(B-V) )
\quad .
\nonumber
\end{eqnarray}
The eight  constants are different for each MK class.}

In order to give an analytical expression
to Conjecture~1
we first analyze the case in which  we are in presence
of calibrated  physical parameters for stars of
the various MK spectral types , see Section~\ref{sec_calibrated}
 and then the   case
of absence of calibration tables,
see Section~\ref{sec_noncalibrated}~.
Different astrophysical environments such as the first 10~pc
and 50~pc , the open clusters and distance determination
of the open clusters are presented in Section~\ref{applications}.
The theoretical dependence by the temperature 
for colors and  bolometric corrections 
are analyzed in Section~\ref{theoretical}.

\section{Presence of calibrated physical parameters}
\label{sec_calibrated}
The $M_{V}$   ,visual magnitude,  against $(B-V)$
can be found star\-ting from five   equations , four of them
were already described in  \cite{zaninetti05}.
When the numerical value of  the symbols  is omitted
the interested reader is demanded to \cite{zaninetti05}.
The luminosity of the star is
\begin{equation}
\log_{10}(\frac {L}{L_{\sun}})  =  0.4 (M_{\mathrm{bol},\sun} -M_{\mathrm{bol}})
\quad,
\end {equation}
where $M_{\mathrm{bol},\sun}$ is the bolometric luminosity of the sun
that according to~\cite{cox} is 4.74.
The  equation
that regulates the total luminosity of a star with it's mass is
\begin{equation}
\log_{10}(\frac {L}{L_{\sun}})  = a_{\mathrm {LM}} +b_{\mathrm {LM}}\log_{10}(\frac {{\mathcal M}}
 {{\mathcal M}_{\sun}})
\label{eqn_lm}
\quad,
\end{equation}
here $L$      is the total luminosity of a star , ${L_{\sun}}  $
the sun's luminosity, ${\mathcal M}$ the star's mass , ${\mathcal
M}_{\sun}$ the sun's  mass , $a_{\mathrm {LM}}$ and $b_{\mathrm{LM}}$ 
two coefficients that  are reported 
in Table~\ref{coefficients} 
for   MAIN  V  ,  GIANTS III and  SUPERGIANTS I
; more details can be found in   \cite{zaninetti05}.
\begin{table}[h] 
\caption{Table of coefficients
derived from the calibrated data 
(see Table~15.7  in~\cite{cox}) 
through the 
least square method }
\label{coefficients}
\begin{tabular}{|l|c|c|c|}
\hline
  & MAIN, V  & GIANTS, III, & SUPERGIANTS I       \\
                   \hline
$K_{\mathrm{BV}}$       &-0.641 $\pm$ 0.01 &  -0.792 $\pm$ 0.06   & -0.749 $\pm$ 0.01 \\
$T_{\mathrm{BV}}[\mathrm {K}]$       &7360   $\pm$  66  &     8527$\pm$ 257    & 8261   $\pm$ 67   \\
$K_{\mathrm{BC}}$       & 42.74 $\pm$ 0.01 &  44.11  $\pm$ 0.06   & 42.87  $\pm$ 0.01 \\
$T_{\mathrm {BC}}[\mathrm {K}]$       & 31556 $\pm$  66  & 36856   $\pm$ 257    & 31573  $\pm$ 67   \\
$a_{\mathrm {LM}}$       &0.062 $\pm$ 0.04 &  0.32   $\pm$ 0.14    & 1.29   $\pm$ 0.32 \\
$b_{\mathrm {LM}}$       &3.43  $\pm$ 0.06 &  2.79   $\pm$ 0.23    & 2.43   $\pm$ 0.26 \\
\hline
\end{tabular}
\end{table}
We remember that the tables of  calibration of MK spectral types 
unify  SUPERGIANTS Ia  and      SUPERGIANTS Ib   into SUPERGIANTS I,
see Table~15.7  in~\cite{cox} and 
Table 3.1 in \cite{deeming}.
From the theoretical side~\cite{Padmanabhan_II_2001} quotes 
$3 < b_{\mathrm {LM}} < 5  $   ;
the fit on the calibrated values gives 
$2.43 < b_{\mathrm {LM}} < 3.43  $   ,
see Table~\ref{coefficients}.

From a visual
inspection of formula~(\ref{eqn_lm}) is possible to conclude  that
a logarithmic expression for the  mass as function of the
temperature will allows us to continue with formulas easy to deal
with. The following form of the mass-temperature relationship is
therefore chosen
\begin{equation}
\log_{10}(\frac {\mathcal M}{\mathcal M_{\sun}})  = a_{\mathrm {MT}}
+b_{\mathrm {MT}}\log_{10}(\frac {T}
 {T_{\sun}})
\label{eqn_mt}
\quad,
\end{equation}
where $T$ is the star's temperature, ${T_{\sun}}$ the sun's
temperature , $a_{\mathrm {MT}}$ and  $b_{\mathrm {MT}}$ two
coefficients that are reported in  Table~\ref{mtdata}
 when  the masses as function of the temperature
  ( e.g.  Table 3.1 in \cite{deeming})  are processed.
According to~\cite{cox} ${T_{\sun}}=5777~K$.
\begin{table}[h]
\caption{Table of  $a_{\mathrm {MT}}$ and
$b_{\mathrm {MT}}$ with masses given by Table 3.1 in  Bowers \& Deeming (1984).}
\label{mtdata}
\begin{tabular}{|l|c|c|c|c|}
\hline
  & main,V  & giants,III  & supergiants,I     &~      \\
\hline
  & ~  &  ~  &    (B-V)$ >$ 0.76  &  (B-V) $ < $ 0.76   \\
                   \hline
$a_{\mathrm {MT}}$       & -7.6569  
                         &  5.8958  
                         & -3.0497  
                         &  4.1993  
                        \\
$b_{\mathrm {MT}}$       &  2.0102   
                         & -1.4563   
                         & -0.8491   
                         &  1.0599   
                        \\
$\chi^2$                 & 28.67
                         & 3.41
                         &  20739
                         &  18.45
                         \\
\hline
\end{tabular}
\end{table}
Due to the fact that the masses of the SUPERGIANTS present a
minimum at $(B-V)$ $\approx$ 0.7 or T$\approx~ 5700~K$ we have divided
the analysis in two. 
Another useful formula is the bolometric
correction $BC$
\begin {equation}
BC  = M_{\mathrm{bol}}  -M_{\mathrm{V}} =
-\frac{T_{\mathrm{BC}}}{T} - 10~\log_{10}~T + K_{\mathrm {BC}}
\quad, \label{bct}
\end {equation}
where $M_{\mathrm{bol}}$   is the absolute bolometric magnitude,
      $M_{V}$     is the absolute visual     magnitude,
      $T_{\mathrm {BC}}$ and  $K_{\mathrm{BC}}$
      are two parameters that are  derived in Table~\ref{coefficients} .
The bolometric correction is always negative but
in~\cite{Allen1973} the analytical formula
was erroneously reported as always positive.

The fifth equation connects the physical variable   $T$  with the
observed color $(B-V)$ , see for example ~\cite{Allen1973},
\begin{equation}
(B-V)=  K_{\mathrm{BV}}   + \frac {T_{\mathrm{BV}}}{T} \quad,
\label{bvt}
\end {equation}
where
 $K_{\mathrm{BV}}$ and $T_{\mathrm{BV}}$ are two parameters that are
derived in Table~\ref{coefficients}  
from the least square fitting procedure.
Conversely in Section~\ref{theoretical}  we will 
explore a series development for  $(B-V)$ as given 
by a Taylor series   in the variable $1/T$.
Inserting formulas~(\ref{bct}) and (\ref{eqn_mt}) in
(\ref{eqn_lm})
 we
obtain
\begin{eqnarray}
M_V=
 - 2.5\,{\it a_{\mathrm {LM}}}- 2.5\,{\it b_{\mathrm {LM}}}\,{\it a_{\mathrm {MT}}}- 
   \nonumber\\
2.5\,{\it
b_{\mathrm {LM}}}\,{\it b_{\mathrm {MT}}} \,{\it log_{10}T}   \nonumber \\
-{\it K_{\mathrm {BC}}}+10\,{\it log_{10}T}+{\frac {{\it
T_{\mathrm {BC}}}}{T}}+ M_{\mathrm{bol},\sun}
 \quad .  \label{eqn_mvt}
\end{eqnarray}
 Inserting equation~(\ref{bvt}) in
(\ref{eqn_mvt}) the following relationship that regulates $M_V$
and $(B-V)$ in the H-R diagram is obtained
\begin{eqnarray}
M_V=
 - 2.5\,{\it a_{\mathrm {LM}}}- 2.5\,{\it b_{\mathrm {LM}}}\,{\it a_{\mathrm {MT}}}-
\nonumber \\
 2.5\,{\it
b_{\mathrm {LM}}}\,{\it b_{\mathrm {MT}}} \,{\it log_{10}(\frac{T_{\mathrm{BV}}}{(B-V) - K_{\mathrm{BV}}})}   
\nonumber \\
-{\it K_{\mathrm {BC}}}+10\,{\it
log_{10}(\frac{T_{\mathrm{BV}}}{(B-V) - K_{\mathrm{BV}}})}+
\nonumber \\
{\frac
{{\it T_{\mathrm {BC}}}}{T_{\mathrm{BV}}}}
 \left [{(B-V) -K_{\mathrm{BV}}} \right] + M_{\mathrm{bol},\sun}
 \quad . \label{eqn_mvbv}
\end{eqnarray}

Up to now  the parameters
$a_{\mathrm {MT}}$ and  $b_{\mathrm {MT}}$  are deduced
from Table 3.1 in \cite{deeming}
and Table~\ref{mtdata}  reports
the  merit function $\chi^2$
computed as
\begin{equation}
\chi^2 =
\sum_{j=1}^n  ( M_V   - M_V^{cal})^2
\quad  ,
\label{chisquare}
\end{equation}
where $M_V^{cal}$ represents the calibration value
for the three MK classes as given
by Table~15.7 in ~\cite{cox}.
From a visual inspection of the $\chi^2$ reported in
Table~\ref{mtdata} we deduced that  different coefficients
of the mass-temperature relationship~(\ref{eqn_mt})
may give better results.
We therefore found by a numerical analysis the
values $a_{\mathrm {MT}}$ and  $b_{\mathrm {MT}}$
that minimizes equation~(\ref{chisquare}) when
$(B-V)$ and  {$M_{\mathrm V}$  are given by the calibrated values
of  Table~15.7 in ~\cite{cox}.
\begin{table}[h]
\caption{Table of  $a_{\mathrm {MT}}$ and $b_{\mathrm {MT}}$ when
$M_V^{cal}$ is given by  calibrated data.} \label{mtdtata_cali}
\begin{tabular}{|l|c|c|c|c|}
\hline
  & main,V  & giants, III, & supergiants, I     &~      \\
\hline
  & ~  &  ~  &    (B-V)$>$0.76  &  (B-V)$<$0.76   \\
                   \hline
$a_{\mathrm {MT}}$       & -7.76  
                         &  3.41  
                         &  3.73   
                         &  0.20   
                        \\
$b_{\mathrm {MT}}$       &  2.06    
                         & -2.68    
                         & -0.64    
                         &  0.24    
                        \\
$\chi^2$                 & 11.86
                         & 0.152
                         & 0.068
                         & 0.567
                         \\
\hline
\end{tabular}
\end{table}
This method to evaluate $a_{\mathrm {MT}}$ and  $b_{\mathrm {MT}}$
is new and allows to compute them in absence of
other ways to deduce the mass of a star.
The absolute visual magnitude with the data
of Table~\ref{mtdtata_cali} is
\begin{eqnarray}
M_{\mathrm V} =
31.34- 
\nonumber \\
3.365\,\ln  \left(  7361.0\, \left( {\it (B-V)}+ 0.6412 \right) ^{-1}
\right) +
\nonumber \\
 4.287\,{\it (B-V)}    \\
 MAIN~SEQUENCE~,~V~~ when \nonumber \\
 ~-0.33<(B-V)<1.64,  \nonumber          
\label{mvmain}
\end{eqnarray}
\begin{eqnarray}
M_{\mathrm V} =
- 109.6+ \nonumber \\
12.51\,\ln  \left(  8528.0\, \left( {\it (B-V)}+ 0.7920 \right)
^{-1} \right) \nonumber \\ +
 4.322\,{\it (B-V)}
\\
GIANTS~,~ III   \quad   0.86  < (B-V) < 1.33
\quad  ,          \nonumber
\label{mvgiants}
\end{eqnarray}

\begin{eqnarray}
M_{\mathrm V} =
- 39.74+ \nonumber \\
 3.565\,\ln  \left(  8261.0\, \left( {\it (B-V)}+ 0.7491 \right)
^{-1} \right) + \nonumber \\
3.822\,{\it (B-V)}   \\
SUPERGIANTS,I~when \nonumber \\
~-0.27<(B-V)<0.76\,,          \nonumber
\label{mvsupergiantsbiggert}
\end{eqnarray}

\begin{eqnarray}
M_{\mathrm V} =
- 61.26+   \nonumber\\
6.050\,\ln  \left(  8261.0\, \left( {\it (B-V)}+ 0.7491 \right)
^{-1} \right) + \nonumber\\
3.822\,{\it (B-V)}       \\
SUPERGIANTS~,~ I \quad when~ \nonumber \\
   0.76    < (B-V) < 1.80
\quad  .          \nonumber
\label{mvsupergiantalowert}
\end{eqnarray}

Is now possible  to build the calibrated and theoretical H-R diagram
, see  Figure~\ref{f01}.
\begin{figure}
\begin{center}
\includegraphics[width=10cm]{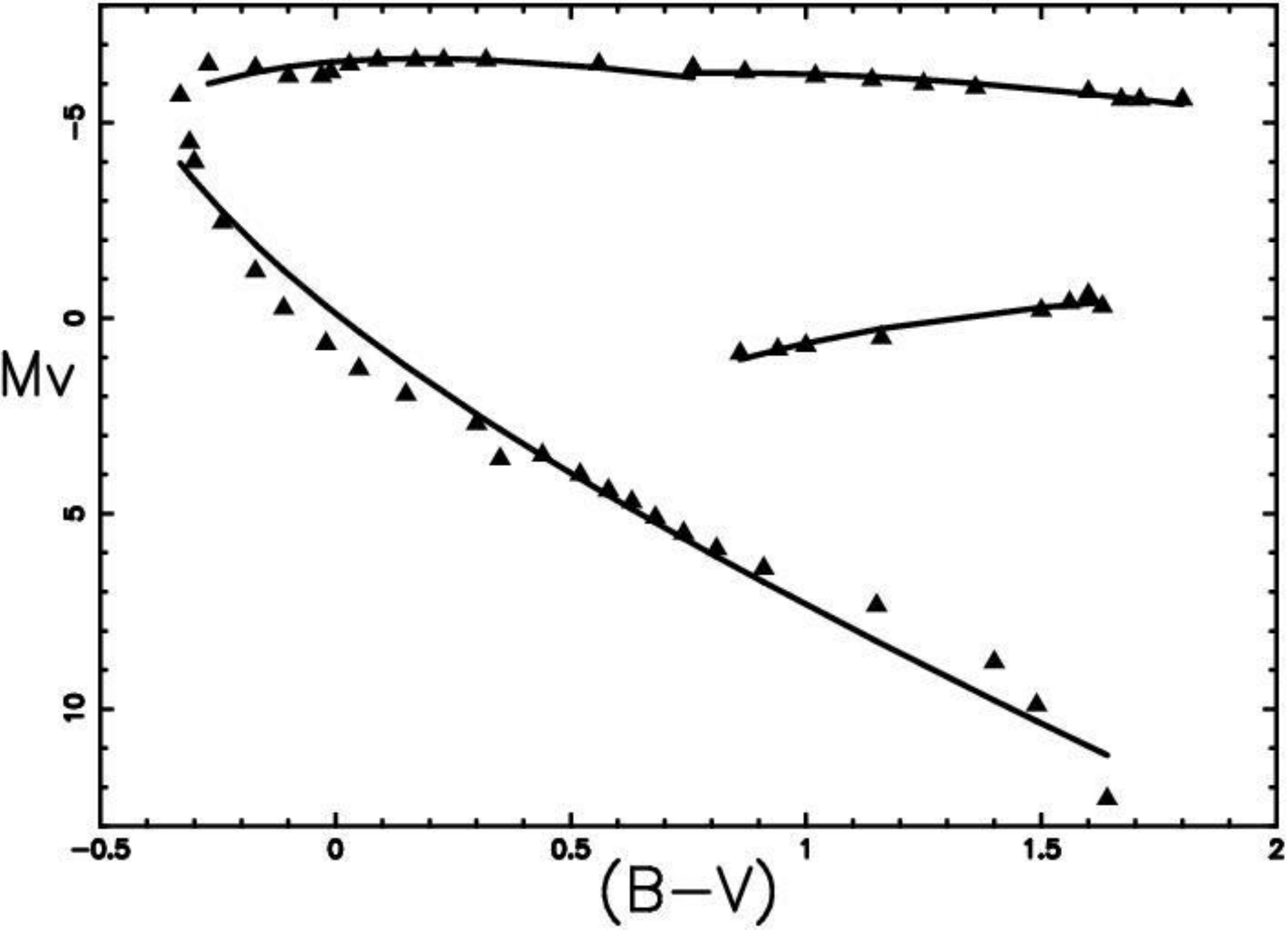}
\end{center}
\caption{$M_{\mathrm V}$  against $(B-V)$
  for calibrated MK stars  (triangles) and  theoretical
  relationship with $a_{\mathrm {MT}}$ and  $b_{\mathrm {MT}}$
  as given by Table~\ref{mtdtata_cali}.
  }
\label{f01}
\end{figure}

\subsection{The mass (B-V) relationship}

Is now possible  to deduce the numerical relationship
that connects the mass of  the star ,${\mathcal M}$,
with  the variable  $(B-V)$
and
the constants
 $a_{\mathrm {MT}}$ and  $b_{\mathrm {MT}}$
\begin{eqnarray}
\label{eqn_mass}
\log_{10}
 ( \frac {{\mathcal M}} {\mathcal M}_{\sun})  =
{\it a_{MT}}+\nonumber \\
{\it b_{MT}}\,\ln  \left( {\frac {{\it T_{BV}}}{{\it (B-V)}-{\it
K_{BV}}
}} \right)  \left( \ln  \left( 10 \right)  \right) ^{-1}
\quad .
\end  {eqnarray}
When
 $a_{\mathrm {MT}}$ and  $b_{\mathrm {MT}}$
are given  by Table~\ref{mtdtata_cali} and the other coefficients
are  as reported in Table~\ref{coefficients}  the following expression
for the mass is  obtained:
\begin{eqnarray}
\label{masses1}
\log_{10}
 ( \frac {{\mathcal M}} {\mathcal M}_{\sun})  =
 7.769 + \nonumber \\
+ 0.8972\,\ln  \left(  7360.9\, \left( {\it (B-V)}+ 0.6411
\right) ^{-1} \right)
\\
MAIN~SEQUENCE~,~ V   ~when \nonumber \\
  -0.33 < (B-V) < 1.64
\quad  ,
\nonumber
\end{eqnarray}

\begin{eqnarray}
\label{masses2}
\log_{10}
 ( \frac {{\mathcal M}} {\mathcal M}_{\sun})  =
10.41  +\nonumber \\ - 1.167\,\ln  \left(  8527.5\, \left( {\it (B-V)}+ 0.792
\right) ^{-1} \right)
\\
GIANTS~,~ III   \quad   0.86  < (B-V) < 1.33
\quad  ,          \nonumber
\end{eqnarray}

\begin{eqnarray}
\label{masses3}
\log_{10}
 ( \frac {{\mathcal M}} {\mathcal M}_{\sun})  =
 0.2  + \nonumber \\ + 0.1276\,\ln  \left(  8261\, \left( {\it (B-V)}+
 0.7491 \right) ^{-1} \right)
\\
SUPERGIANTS~,~ I \quad ~when~ \nonumber \\
  -0.27   < (B-V) < 0.76
\quad  ,          \nonumber
\end{eqnarray}

\begin{eqnarray}
\label{masses4}
\log_{10}
 ( \frac {{\mathcal M}} {\mathcal M}_{\sun})  =
3.73  +\nonumber \\   - 0.2801\,\ln  \left(  8261\, \left( {\it (B-V)}+ 0.7491
\right) ^{-1} \right)
\\
SUPERGIANTS~,~ I \quad when~ \nonumber \\  0.76    < (B-V) < 1.80
\quad  .          \nonumber
\end{eqnarray}
Figure~\ref{f02} reports the logarithm of  the mass
as function of $(B-V)$  for the three classes here considered
(points) as well the theoretical relationships given by
equations~(\ref{masses1}-\ref{masses4})
 (full lines).

\begin{figure}
\begin{center}
\includegraphics[width=10cm]{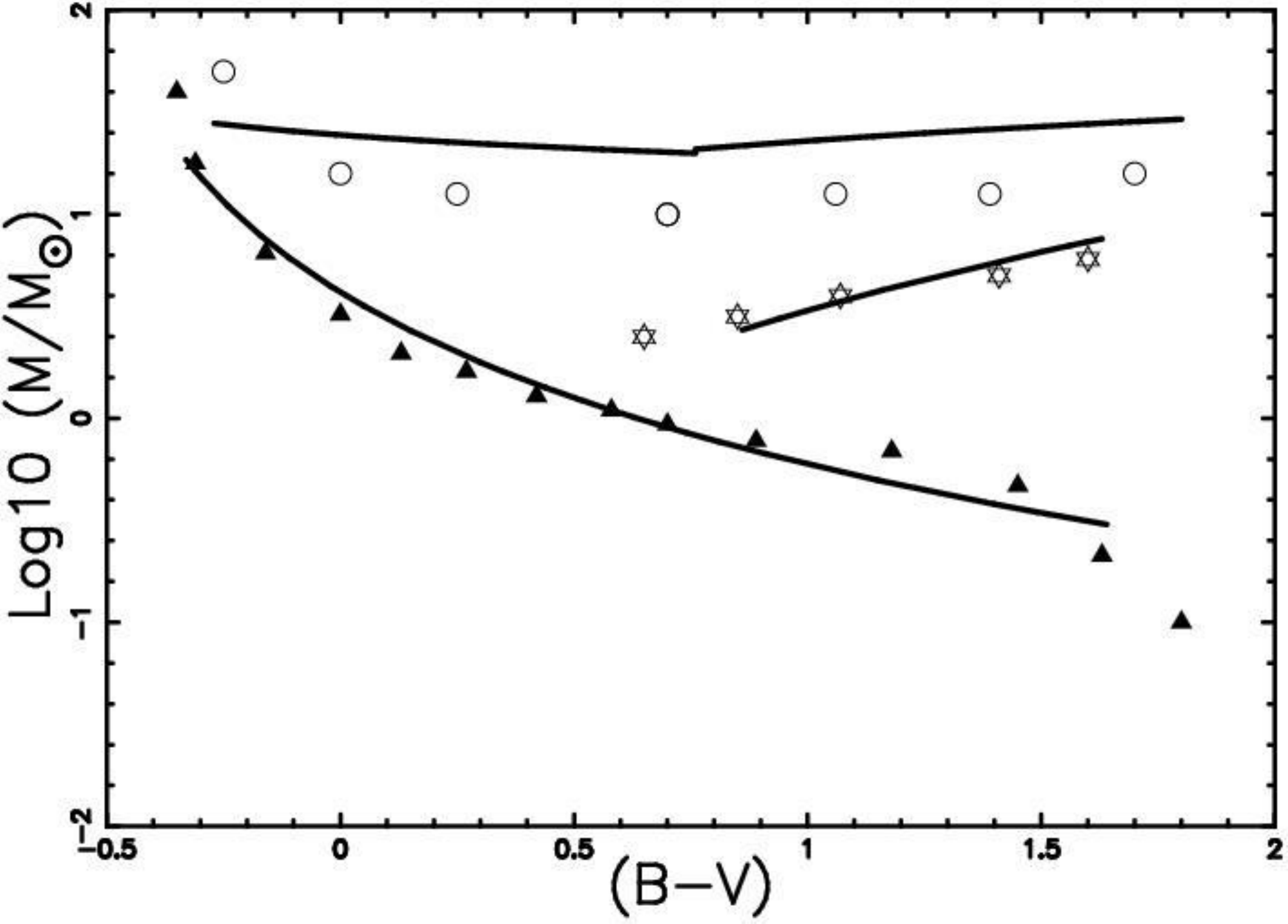}
\end{center}
\caption{$\log_{10}(\frac{{\mathcal{M}}} {{\mathcal{M}_{\sun}}}$)
  against $(B-V)$
  for calibrated MK stars :
  MAIN~SEQUENCE~,~ V    (triangles)   ,
  GIANTS~,~ III         (empty stars)
  and SUPERGIANTS~,~ I  (empty circles).
  The theoretical relationship as
  given by formulas~(\ref{masses1}-\ref{masses4})
  is reported as a full line.
  }
\label{f02}
\end{figure}

\subsection{The radius $(B-V)$ relationship}

The radius of a star can be found from the
Stefan-Boltzmann law, see for example formula~(5.123) in~\cite{lang} .
In our framework the radius is
\begin{eqnarray}
\label{eqn_radius}
\log_{10}
 ( \frac {R}{R{\sun}})  =  \nonumber \\
1/2\,{\it a_{LM}}+1/2\,{\it b_{LM}}\,{\it a_{MT}}+
2\,{\frac {\ln  \left( {\it
T_{\sun}} \right) }{\ln  \left( 10 \right) }}
+\nonumber\\
+1/2\,{\it b_{LM}}\,{\it b_{MT}}\,
\ln  \left( {\frac {{\it T_{BV}}}{{\it {(B-V)}}-{\it K_{BV}}}} \right)  \left(
\ln  \left( 10 \right)  \right) ^{-1}
\nonumber\\
-2\,\ln  \left( {\frac {{\it T_{BV}}
}{{\it {(B-V)}}-{\it K_{BV}}}} \right)  \left( \ln  \left( 10 \right)
 \right) ^{-1}
\quad .
\end  {eqnarray}
When  the  coefficients are given  by Table~\ref{mtdtata_cali} and
Table~\ref{coefficients}  the radius is
\begin{eqnarray}
\log_{10}
 ( \frac {R}{R{\sun}})  =
- 5.793+ \nonumber \\
0.6729\,\ln  \left(  7360\, \left( {\it (B-V)}+
 0.6411 \right) ^{-1} \right)
\\
MAIN~SEQUENCE~,~ V  ~when \nonumber \\  -0.33 < (B-V) < 1.64
\quad ,
\nonumber
\label{raggio1}
\end  {eqnarray}

\begin{eqnarray}
\log_{10}
 ( \frac {R}{R{\sun}})  =
22.25   \nonumber \\
-   2.502\,\ln  \left(  8527\, \left( {\it (B-V)}+ 0.7920
\right) ^{-1} \right)
\\
GIANTS~,~ III   ~ when \nonumber \\
   0.86  < (B-V) < 1.33
\quad ,
\nonumber
\label{raggio2}
\end  {eqnarray}

\begin{eqnarray}
\log_{10}
 ( \frac {R}{R{\sun}})  =
 8.417 - \nonumber \\
0.7129\,\ln  \left(  8261\, \left( {\it (B-V)}+
 0.7491 \right) ^{-1} \right)
   \\
SUPERGIANTS~,~ I \quad   ~when \nonumber \\
-0.27   < (B-V) < 0.76
\quad  ,
\nonumber
\label{raggio3}
\end  {eqnarray}

\begin{eqnarray}
\log_{10}
 ( \frac {R}{R{\sun}})  =
 12.71      \nonumber\\
- 1.21\,\ln  \left(  8261\, \left( {\it (B-V)}+
 0.7491 \right) ^{-1} \right)
 \\
SUPERGIANTS~,~ I \quad when~\nonumber \\ \quad   0.76    < (B-V) < 1.80
\quad  .          \nonumber
\label{raggio4}
\end  {eqnarray}
Figure~\ref{f03} reports the radius
as function of $(B-V)$  for the three classes  (points)
as well the theoretical
relationship given by equations~(\ref{raggio1}-\ref{raggio4})
 (full lines).

\begin{figure}
\begin{center}
\includegraphics[width=10cm]{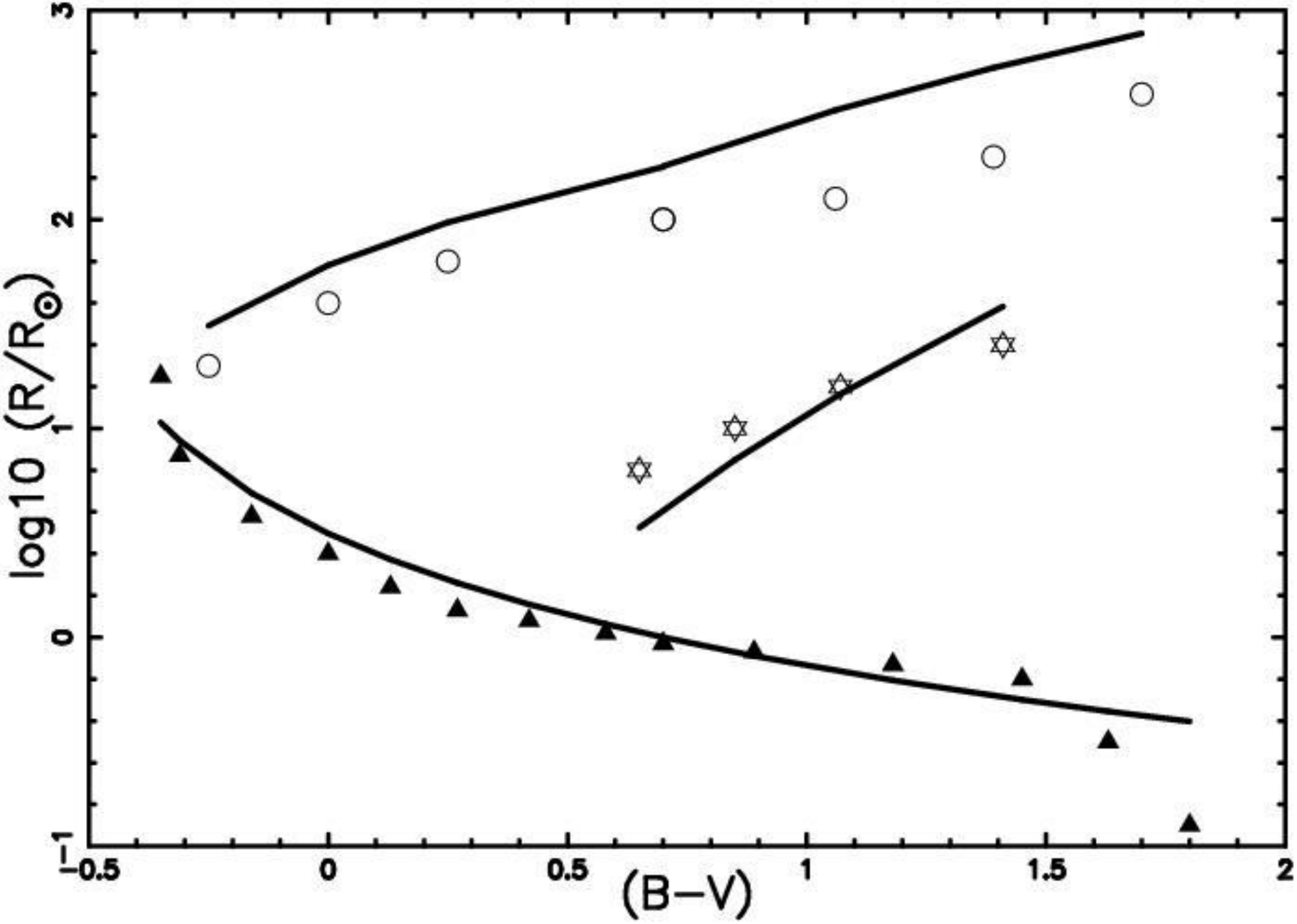}
\end{center}
\caption{$\log_{10}
 ( \frac {R}{R{\sun}}) $)
  against $(B-V)$
  for calibrated MK stars :
  MAIN~SEQUENCE~,~ V   (triangles) ,
  GIANTS~,~ III        (empty stars)
  and SUPERGIANTS~,~ I (empty circles).
  The theoretical relationships as
  given by formulas~(\ref{raggio1}-\ref{raggio4})
  are reported as a full line.
  }
\label{f03}
\end{figure}

\subsection{The luminosity  $(B-V)$ relationship}

The luminosity of a star can be parametrized as
\begin{eqnarray}
\label{eqn_luminosity}
\log_{10}
 ( \frac {L}{L{\sun}})  =
{\it a_{LM}} +{\it b_{LM}} {\it a_{MT}}    \nonumber \\
+{\it b_{LM}}\left ( {\it b_{MT}}\,\ln  \left( {\frac {
{\it T_{BV}}}{{\it (B-V)}-{\it K_{BV}}}} \right)  \frac{1}{\ln  \left( 10
 \right)  }  \right) \, .
\end  {eqnarray}
When  the  coefficients are given  by Table~\ref{mtdtata_cali} and
Table~\ref{coefficients}  the luminosity is
\begin{eqnarray}
\log_{10}
 ( \frac {L}{L{\sun}})  =
- 26.63+   \nonumber \\
+ 3.083\,\ln  \left(  7360.9\, \left( {\it (B-V)}+
 0.6411 \right) ^{-1} \right)
\\
MAIN~SEQUENCE~,~ V   ~when \nonumber \\
   -0.33 < (B-V) < 1.64
\quad ,
\nonumber
\label{luminosita1}
\end  {eqnarray}

\begin{eqnarray}
\log_{10}
 ( \frac {L}{L{\sun}})  =
29.469 +
\nonumber \\
 -3.2676\,\ln  \left(  8527.59\, \left( {\it (B-V)}+ 0.7920
\right) ^{-1} \right)
\\
GIANTS~,~ III   \quad   0.86  < (B-V) < 1.33
\quad ,
\nonumber
\label{luminosita2}
\end  {eqnarray}

\begin{eqnarray}
\log_{10}
 ( \frac {L}{L{\sun}})  =
 1.7881+
\nonumber \\
+ 0.3112\,\ln  \left(  8261.19\, \left( {\it (B-V)}+
 0.7491 \right) ^{-1} \right)
  \\
SUPERGIANTS~,~ I  ~when~ \nonumber \\  -0.27   < (B-V) < 0.76
\quad  ,
\nonumber
\label{luminosita3}
\end  {eqnarray}

\begin{eqnarray}
\log_{10}
 ( \frac {L}{L{\sun}})  =
10.392+
\nonumber \\
- 0.682\,\ln  \left(  8261.19\, \left( {\it (B-V)}+ 0.749
\right) ^{-1} \right)
\nonumber \\
SUPERGIANTS~,~ I \quad when~ \nonumber \\   0.76    < (B-V) < 1.80
\quad  .
\label{luminosita4}
\end  {eqnarray}
Figure~\ref{f04} reports the luminosity as function
of $(B-V)$  for the three classes  (points) as well as the
theoretical relationship given by
equations~(\ref{luminosita1}-\ref{luminosita4})
 (full lines).

\begin{figure}
\begin{center}
\includegraphics[width=10cm]{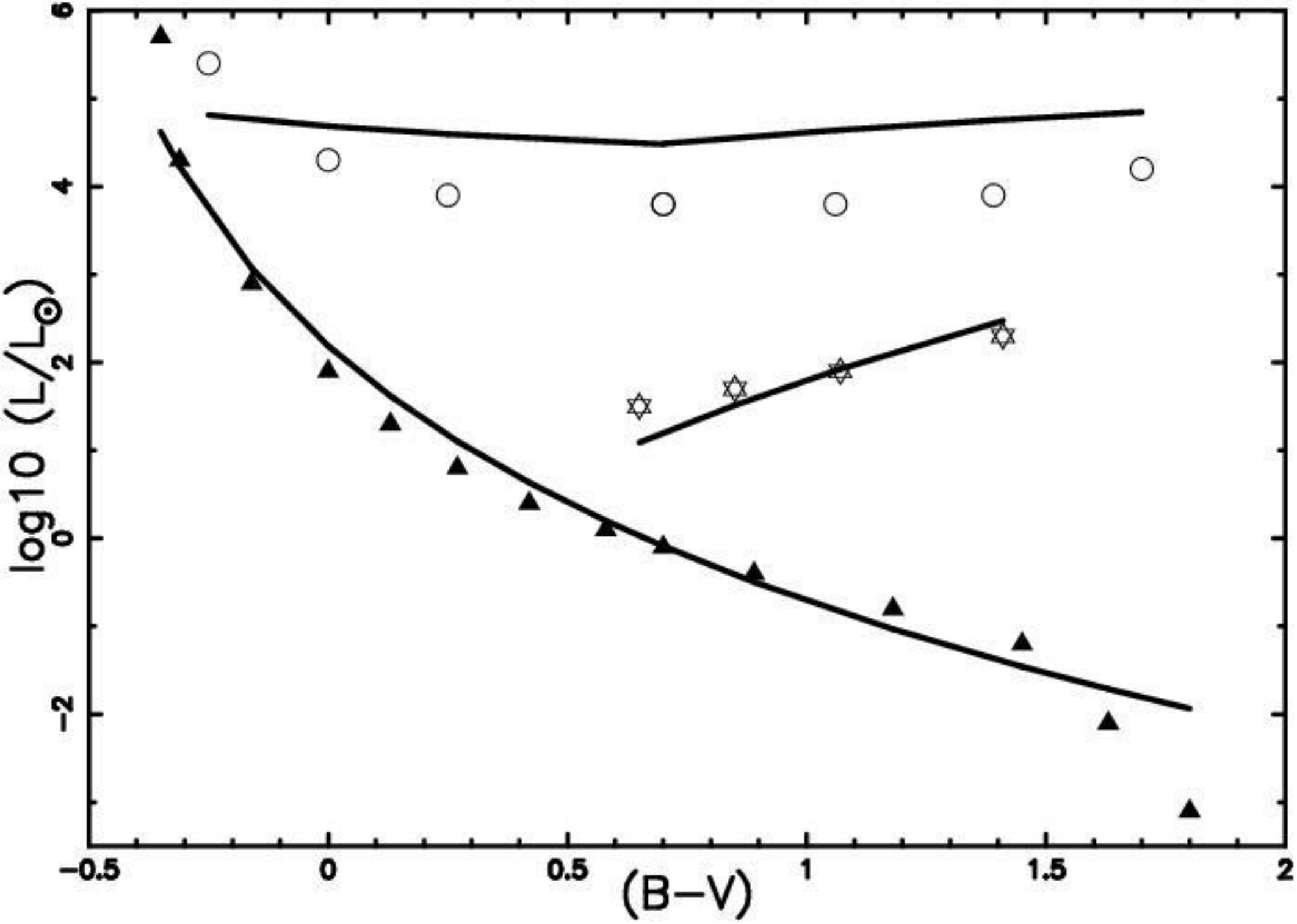}
\end{center}
\caption{$\log_{10}
 ( \frac {L}{L{\sun}}) $
  against $(B-V)$
  for calibrated MK stars :
  MAIN~SEQUENCE~,~ V   (triangles) ,
  GIANTS~,~ III        (empty stars)
  and SUPERGIANTS~,~ I (empty circles).
  The theoretical relationship as given by
  formulas~(\ref{luminosita1}-\ref{luminosita4})
  is reported as a full line.
  }
\label{f04}
\end{figure}
\section{Absence of calibrated physical parameters}
\label{sec_noncalibrated} The already derived framework can be
applied to a class of stars  in which the calibration data of $BC$
, $(B-V)$ , $\mathcal {M}$ , $L$ versus the temperature are
absent, for example the white dwarfs. The presence of the fourth
edition of the Villanova Catalog of Spectroscopically Identified
White Dwarfs, see \cite{McCook1999}, makes possible to build the
H-R diagram of 568 white dwarfs that have trigonometric parallax.
Once the  observed absolute magnitude is derived the  merit
function $\chi^2$ is computed as
\begin{equation}
\chi^2 =
\sum_{j=1}^n  ( M_V   - M_V^{obs})^2
\quad  ,
\label{chisquaretheo}
\end{equation}
where $M_V^{obs}$ represents
the observed value of the absolute magnitude and
the theoretical absolute magnitude, $ M_V $ , is given
by equation~(\ref{eqn_mvbv}).
The four parameters
$K_{\mathrm{BV}}$ ,
$T_{\mathrm{BV}}$ ,
$T_{\mathrm {BC}}$ and  $K_{\mathrm{BC}}$
are supplied by theoretical arguments, i.e.
numerical integration of the fluxes as given
by the Planck distribution ,see~\cite{Planck_1901}.
The remaining four unknown parameters
$a_{\mathrm {MT}}$ , $b_{\mathrm {MT}}$  ,
$a_{\mathrm {LM}}$ and
$b_{\mathrm {LM}}$ are supplied by the minimization
of equation~(\ref{chisquaretheo}).
Table~\ref{coefficients_dwarf} reports the eight parameters
that allow to build the H-R diagram.

\begin{table}[h]
\caption{Table of the adopted coefficients for white dwarfs.}
\label{coefficients_dwarf}
\begin{tabular}{|l|c|c|}
\hline
Coefficient  & value   &  method         \\
                   \hline
$K_{\mathrm{BV}}$              & -0.4693058 &  from~the~Planck~law    \\
$T_{\mathrm{BV}}[\mathrm {K}]$ & 6466.229   &  from~the~Planck~law   \\
$K_{\mathrm{BC}}$              & 42.61225   &  from~the~Planck~law   \\
$T_{\mathrm {BC}}[\mathrm {K}]$& 29154.75   &  from~the~Planck~law   \\
$a_{\mathrm {LM}}$             & 0.28      &  minimum~$\chi^2$~on~real~data \\
$b_{\mathrm {LM}}$             & 2.29    &  minimum~$\chi^2$~on~real~data    \\
$a_{\mathrm {MT}}$             & -7.80  &  minimum~$\chi^2$~on~real~data     \\
$b_{\mathrm {MT}}$             & 1.61       &  minimum~$\chi^2$~on~real~data     \\
\hline
\end{tabular}
\end{table}

\begin{figure}
\begin{center}
\includegraphics[width=10cm]{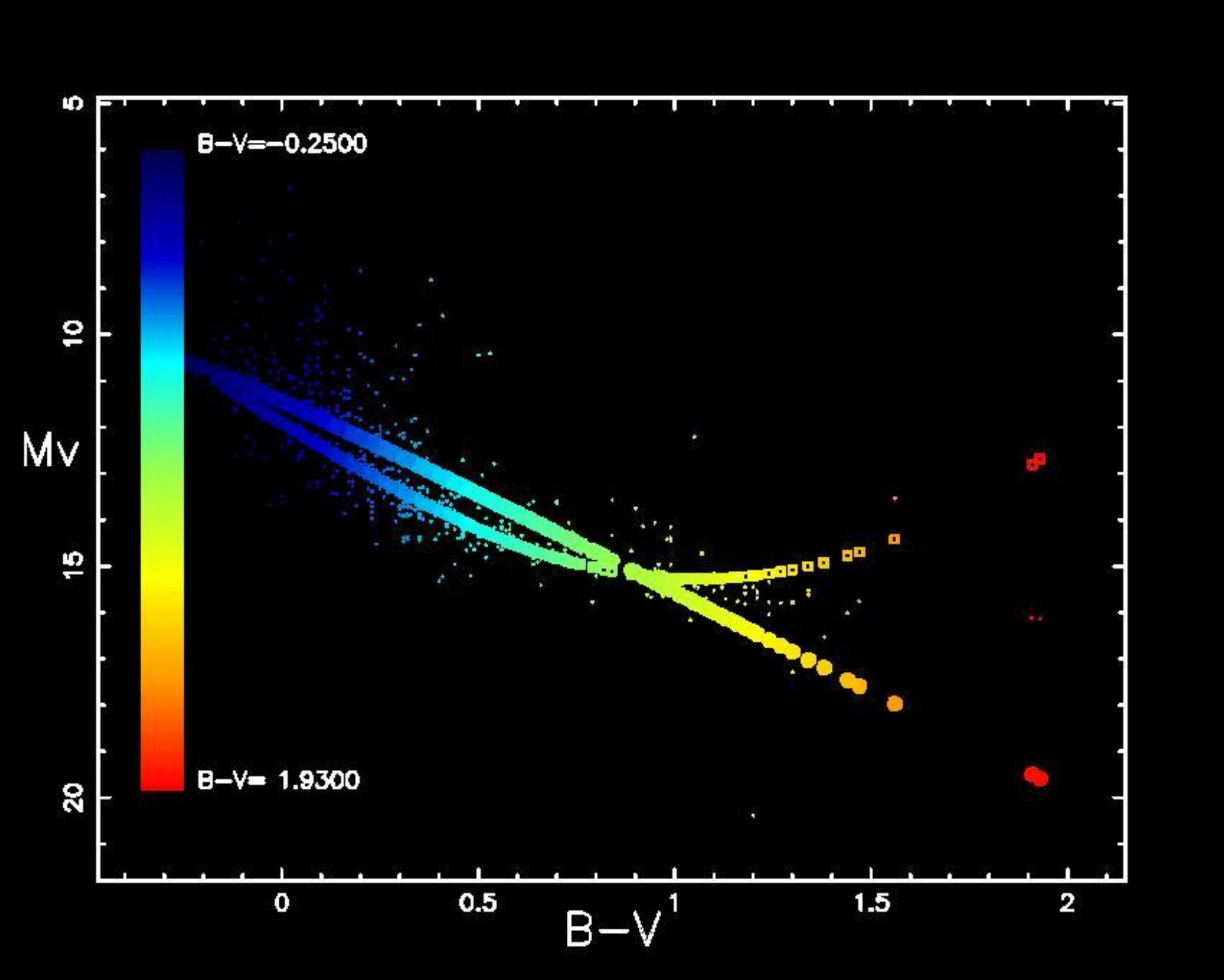}
\end{center}
\caption{$M_{\mathrm V}$  against $(B-V)$   (H-R diagram)
   of the fourth edition of the Villanova Catalog
   of Spectroscopically Identified White Dwarfs.
    The observed stars  are represented through small  points,
    the theoretical relationship of white dwarfs through big full points
    and  the reference relationship given by
    formula~(\ref{reference}) is represented as a little square points.
         }
\label{f05}
\end{figure}
The numerical expressions for the  absolute
magnitude (equation (\ref{eqn_mvbv}))
 , radius (equation (\ref{eqn_radius})) ,
mass (equation (\ref{eqn_mass})) and
luminosity~(equation~(\ref{eqn_luminosity}))  are:
\begin{eqnarray}
M_{\mathrm V} =
8.199   \nonumber \\
+ 0.3399\,\ln  \left(  6466.0\, \left( {\it (B-V)}+ 0.4693 \right) ^{-1}
\right) + \nonumber \\
 4.509\,{\it (B-V)}
\\
white~dwarf,   \quad   -0.25 < (B-V) < 1.88
\quad  ,          \nonumber
\label{mvdwarf}
\end{eqnarray}
\begin{eqnarray}
\label{radiidwarf}
\log_{10}
 ( \frac {R} { R_{\sun}})  =
- 1.267 \nonumber\\
 -   0.0679\,\ln  \left(  6466\, \left( {\it (B-V)}+
 0.4693 \right) ^{-1} \right)
\\
white~dwarf,   \quad   -0.25 < (B-V) < 1.88
\quad  ,
\nonumber
\end{eqnarray}

\begin{eqnarray}
\label{massesdwarf}
\log_{10}
 ( \frac {{\mathcal M}} {\mathcal M}_{\sun})  =
- 7.799  \nonumber \\  + 0.6992\,\ln  \left(  6466\, \left( {\it (B-V)}+
 0.4693 \right) ^{-1} \right)
\\
white~dwarf ,   \quad   -0.25 < (B-V) < 1.88
\quad  ,
\nonumber
\end{eqnarray}

\begin{eqnarray}
\label{luminositydwarf}
\log_{10}
 ( \frac {L} { L_{\sun}})  =
- 17.58 + \nonumber \\
1.601 \,\ln  \left(  6466\, \left( {\it (B-V)}+
 0.4693 \right) ^{-1} \right)
\\
white~dwarf,   \quad   -0.25 < (B-V) < 1.88
\quad  .
\nonumber
\end{eqnarray}

Figure~\ref{f05}
reports the observed absolute visual magnitude
of the white dwarfs as well as  the fitting curve;
Table~\ref{whited} reports the minimum , the average
and the maximum of the three derived physical  quantities.
\begin{table}[h]
\caption{Table of  derived physical parameters
of the Villanova Catalog
of Spectroscopically Identified White Dwarfs}
\label{whited}
\begin{tabular}{|l|c|c|c|}
\hline
 parameter  & min   &  average  &  maximum      \\
\hline
$\mathcal {M}/\mathcal{M}_{\sun}$ & $5.28~10^{-3}$  & $4.23~10^{-2}$  & $0.24$  \\
${R}/{R}_{\sun}$ & $1.07~10^{-2}$  & $1.31~10^{-2}$  &  $1.56~10^{-2}$  \\
${L}/{L}_{\sun}$ & $1.16~10^{-5}$  & $2.6~10^{-3}$  &   $7.88~10^{-2}$  \\
\hline
\hline
\end{tabular}
\end{table}

Our results can be compared with the color-magnitude relation
as suggested by~\cite{McCook1999},
where the  color-magnitude calibration due
to~\cite{Dahn1982} is adopted ,
\begin{eqnarray}
M_V=  11.916\, \left( {\it (B-V)}+1 \right) ^{ 0.44}- 0.011
\nonumber  \\
when  (B-V) < 0.4  \nonumber \\
\label{reference}
~~~                \\
M_V=11.43+ 7.25\,{\it (B-V)}- 3.42\,{{\it (B-V)}}^{2}
\nonumber\\
  when ~  0.4< (B-V)
\nonumber
\end{eqnarray}

The previous formula~(\ref{reference}) is  reported
in  Figure~\ref{f05} as a line made by little squares 
and Table~\ref{chi2_dwarf}   reports   
the $\chi^2$   computed as in formula~(\ref{chisquaretheo})
for our formula~(\ref{mvdwarf}) and  for
the reference formula~(\ref{reference}).
From a visual inspection  of Table~\ref{chi2_dwarf}
is possible to conclude that our relationship represents
a better fit of the data in respect to the reference formula.

\begin{table}[h]
\caption{
Table of  $\chi^2$
when the observed data 
are those of the fourth edition of the Villanova Catalog
of Spectroscopically Identified White Dwarfs.
}
\label{chi2_dwarf}
\begin{tabular}{|l|c|}
\hline
 equation   &   $\chi^2$       \\
\hline
our~formula~(\ref{mvdwarf})          & 710            \\
reference~formula~(\ref{reference})  & 745            \\
\hline
\hline
\end{tabular}
\end{table}

The three classic white dwarfs that are
Procyon~B, Sirius B, and 40 Eridani-B
can also  be analyzed when $(B-V)$   is given
by Wikipedia (http://www.wikipedia.org/).
The results are reported in Table~\ref{EridaniB},
                                  \ref{ProcyonB}
                                  and
                                  \ref{SiriusB}
where is also possible to visualize  the
data suggested by Wikipedia.

\begin{table}[h]
\caption{Table of  derived physical parameters of
40~Eridani-B where $(B-V)$=0.04 }
\label{EridaniB}
\begin{tabular}{|l|c|c|}
\hline
 parameter  &   here   &  suggested in Wikipedia       \\
\hline
$M_V$                              &  11.59          & 11.01            \\
$\mathcal {M}/\mathcal{M}_{\sun}$  & $6.41~10^{-2}$  &   $0.5        $  \\
${R}/{R}_{\sun}$                   & $1.23~10^{-2}$  &   $2~10^{-2}  $  \\
${L}/{L}_{\sun}$                   & $3.53~10^{-3}$  &   $3.3~10^{-3}$  \\
\hline
\hline
\end{tabular}
\end{table}

\begin{table}[h]
\caption{Table of  derived physical parameters of  Procyon B where
 $(B-V)$=0.0 }
\label{ProcyonB}
\begin{tabular}{|l|c|c|}
\hline
 parameter  &   here   &  suggested in Wikipedia       \\
\hline
$M_V                            $  & 11.43           &   13.04         \\
$\mathcal {M}/\mathcal{M}_{\sun}$  & $7.31~10^{-2}$  &   $0.6$          \\
${R}/{R}_{\sun}$                   & $1.21~10^{-2}$  &   $2~10^{-2}$    \\
${L}/{L}_{\sun}$                   & $4.77~10^{-3}$  &   $5.5~10^{-4}$  \\
\hline
\hline
\end{tabular}
\end{table}

\begin{table}[h]
\caption{Table of  derived physical parameters of  Sirius B where
 $(B-V)$=-0.03 }
\label{SiriusB}
\begin{tabular}{|l|c|c|}
\hline
 parameter  &   here   &  suggested in Wikipedia       \\
\hline
$M_V                            $  & 11.32           &   11.35         \\
$\mathcal {M}/\mathcal{M}_{\sun}$  & $8.13~10^{-2}$  &   $0.98$          \\
${R}/{R}_{\sun}$                   & $1.21~10^{-2}$  &   $0.8~10^{-2}$    \\
${L}/{L}_{\sun}$                   & $6.08~10^{-3}$  &   $2.4~10^{-3}$  \\
\hline
\hline
\end{tabular}
\end{table}

\section{Application to the astronomical environment}
\label{applications}
The stars in the first 10~pc , as observed by
Hipparcos (\cite{Esa}) , belong to the
MAIN  V and Figure~\ref{f06}  reports
the observed stars, the calibration stars and the
theoretical relationship given by
equation~(\ref{mvmain}) as a continuous line.
\begin{figure}
\begin{center}
\includegraphics[width=10cm]{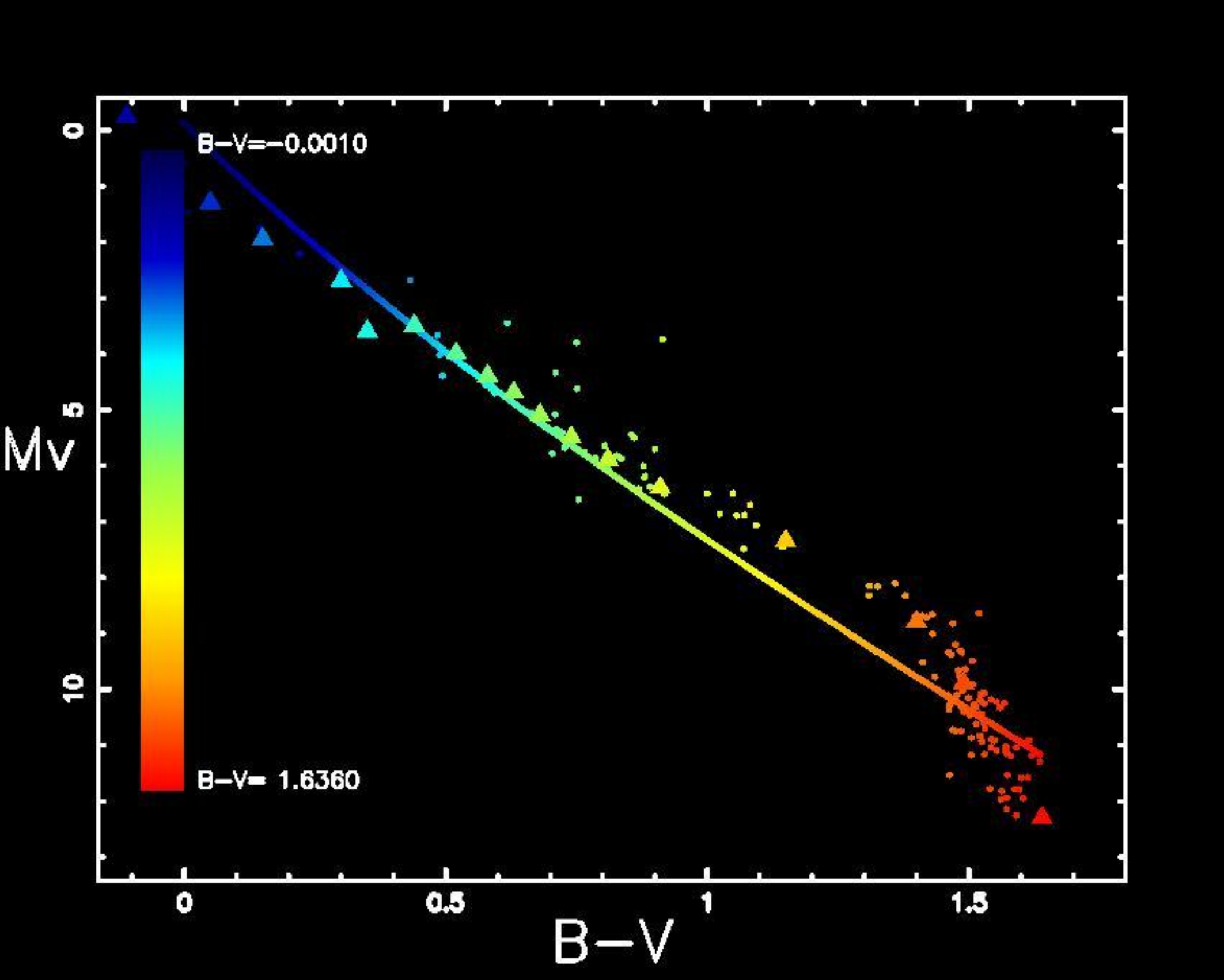}
\end{center}
\caption{$M_{\mathrm V}$  against $(B-V)$   (H-R diagram) in the first  10~pc.
    The observed stars  are represented through  points,
    the calibrated data of MAIN  V with great triangles and
    the theoretical relationship of MAIN  V with a full line.
         }
\label{f06}
\end{figure}

Different is the situation in the first
50~pc where both the MAIN V and the
GIANTS III are present , see
Figure~\ref{f07} where
the theoretical relationship for GIANTS III is
given by  equation~(\ref{mvgiants}).
\begin{figure}
\begin{center}
\includegraphics[width=10cm]{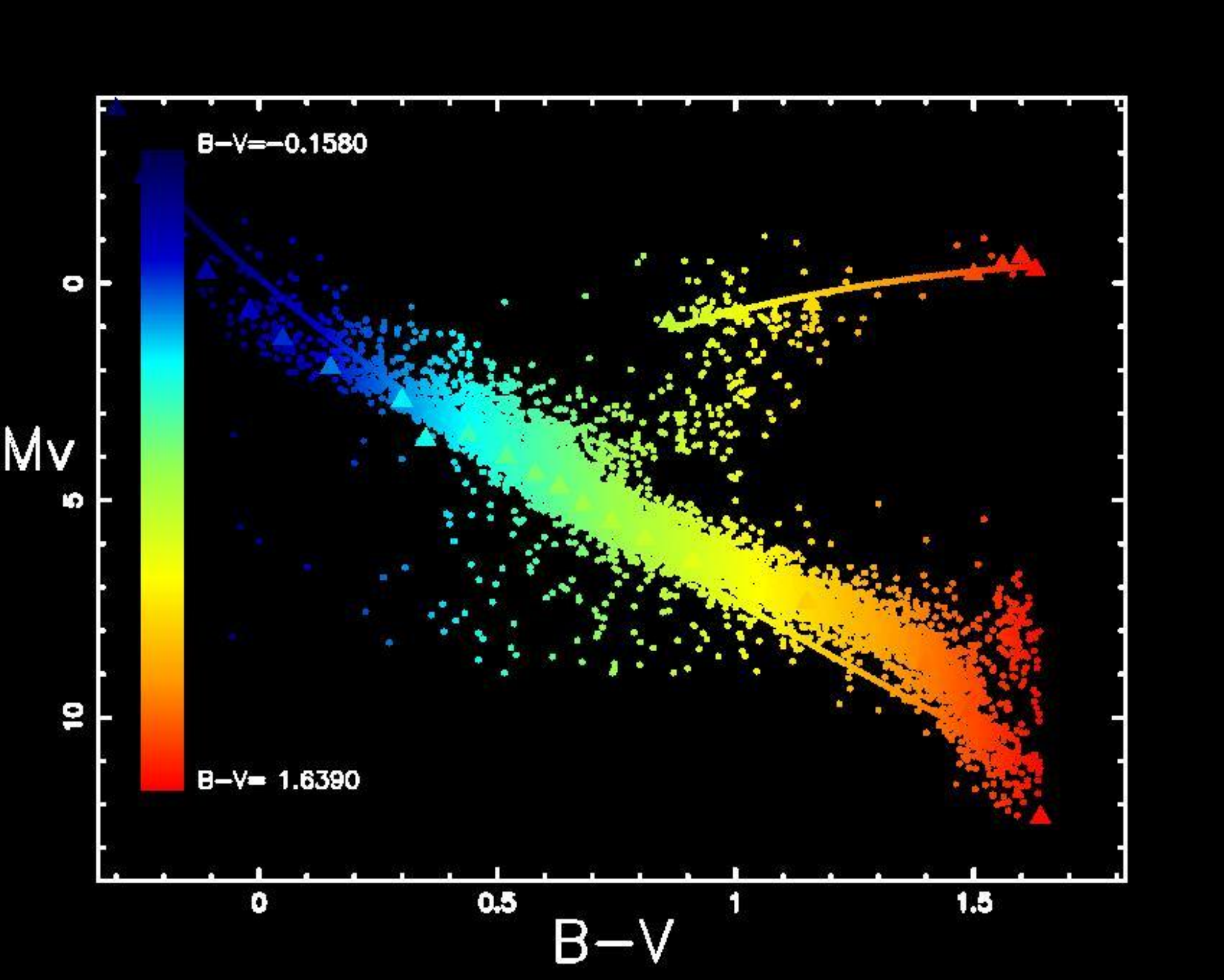}
\end{center}
\caption{$M_{\mathrm V}$  against $(B-V)$   (H-R diagram) in the first  50~pc.
    The observed stars  are represented through  points,
    the calibrated data of MAIN  V and  GIANTS III with great triangles ,
    the theoretical relationship of MAIN  V and
    GIANTS III with a full line.
         }
\label{f07}
\end{figure}

Another astrophysical environment is that of the Hyades cluster,
with $(B-V)$ and $m_v$    as given
in~\cite{Stern1995} and
 available on the
VizieR Online Data Catalog. The H-R
diagram is build in absolute magnitude adopting a distance of
45~$pc$  for the  Hyades, see Figure~\ref{f08} , where also
the theoretical relationship of MAIN  V   is reported.

\begin{figure}
\begin{center}
\includegraphics[width=10cm]{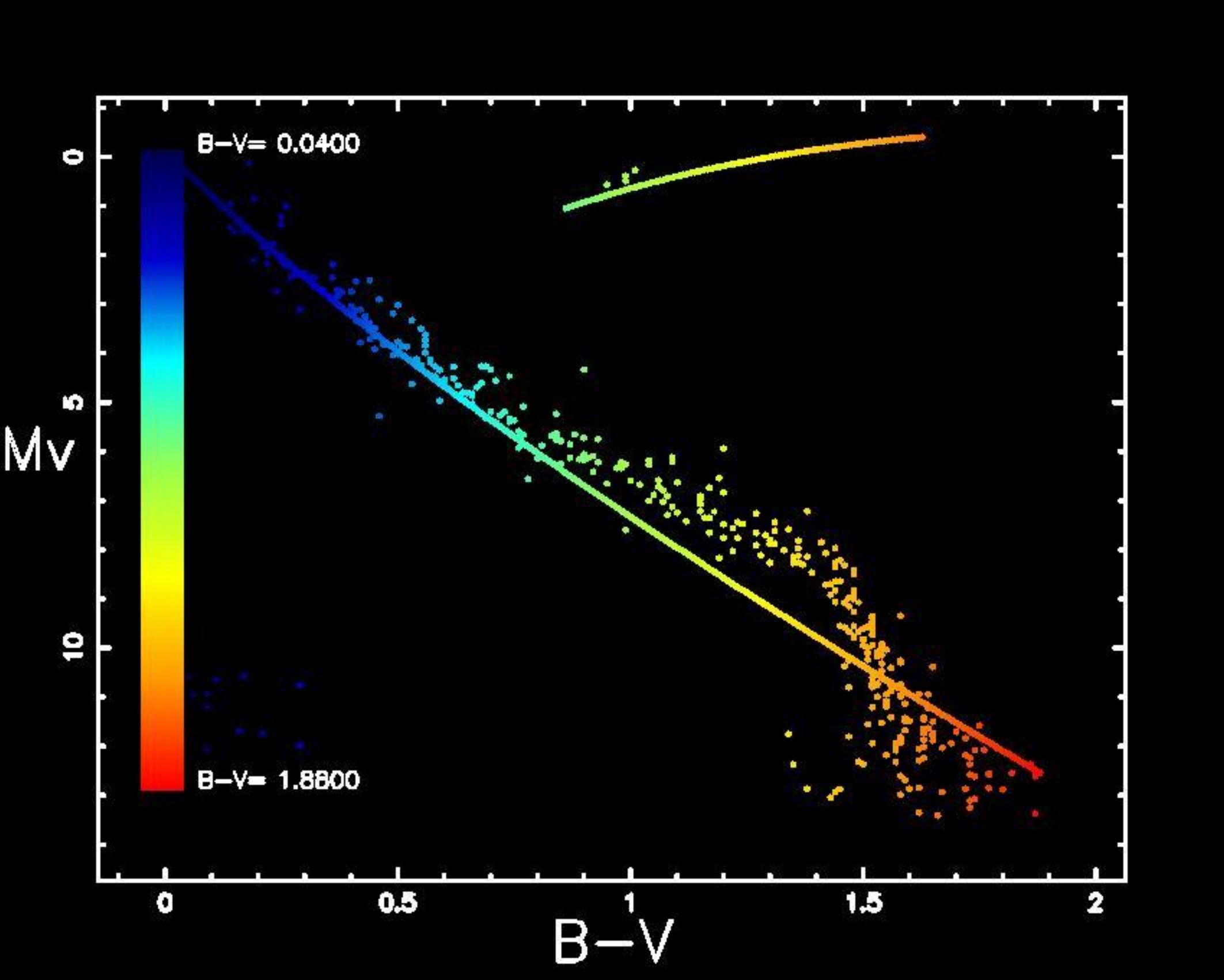}
\end{center}
\caption{$M_{\mathrm V}$  against $(B-V)$   (H-R diagram)
    for the Hyades .
    The observed stars  are represented through  points,
    the theoretical relationship of MAIN  V with a full line.
         }
\label{f08}
\end{figure}

Another  interesting open cluster is that of the Pleiades
with the data  of  \cite{Micela1999} and
 available on the
VizieR Online Data Catalog.

Concerning
the distance of the Pleiades we adopted 135 $pc$ , according
to~\cite{Bouy2006} ; other authors suggest 116 $pc$ as deduced
from  the Hipparcos data , see~\cite{Mermilliod1997}. The H-R
diagram of the Pleiades  is reported in Figure~\ref{f09}.

\begin{figure}
\begin{center}
\includegraphics[width=10cm]{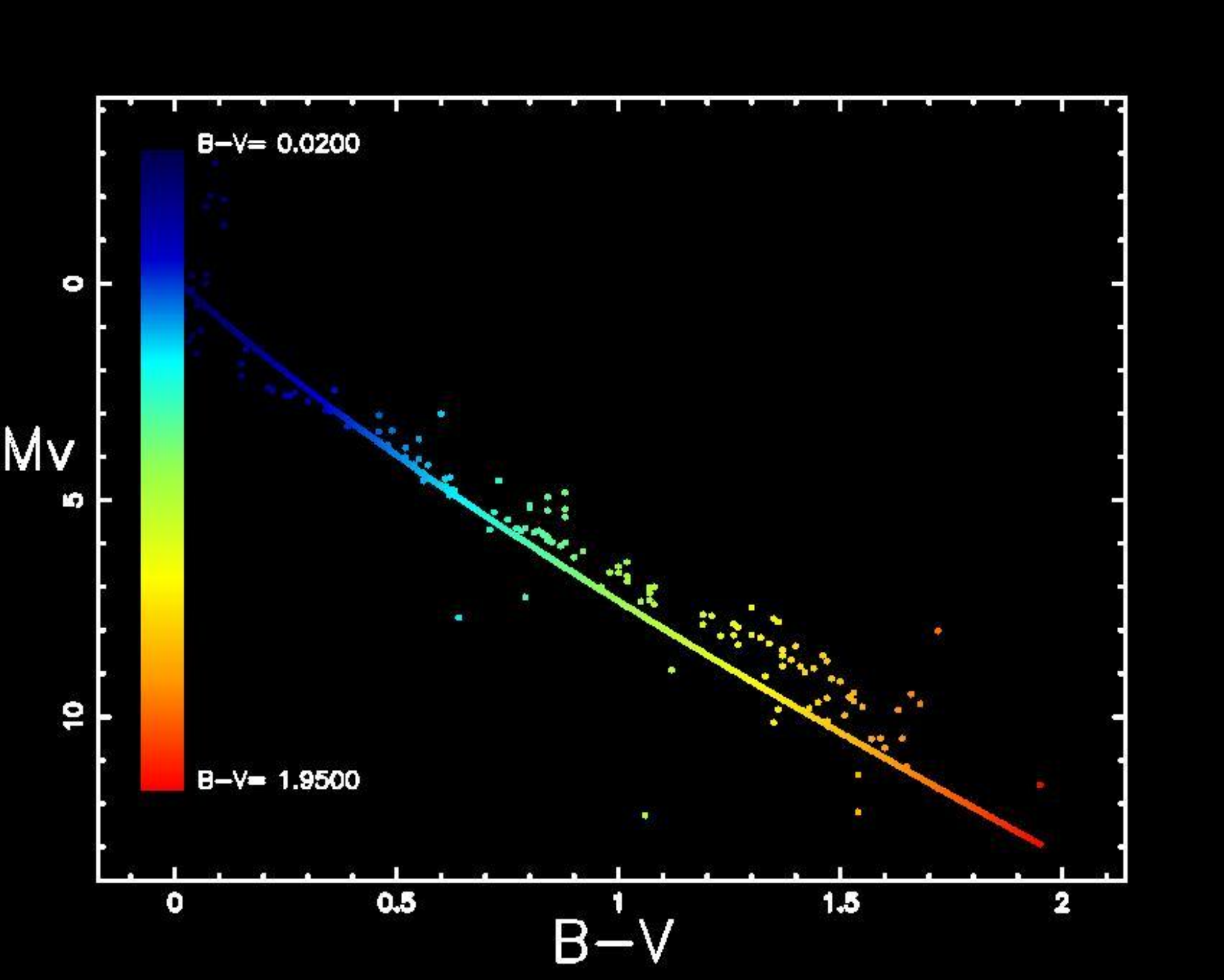}
\end{center}
\caption{$M_{\mathrm V}$  against $(B-V)$   (H-R diagram)
    for the Pleiades.
    The observed stars  are represented through  points,
    the theoretical relationship of MAIN  V with a full line.
         }
\label{f09}
\end{figure}

\subsection{Distance  determination}

The distance of an open  cluster can be found through the
following algorithm :
\begin{enumerate}
\item The absolute magnitude is computed introducing  a guess
      value of the distance.
\item Only the stars belonging  to  MAIN  V are selected.
\item The $\chi^2$   between observed an theoretical
      absolute magnitude ( see formula~(\ref{mvmain}))
      is computed for different distances.
\item The distance of the  open cluster is that connected
      with the value that minimize the $\chi^2$.
\end{enumerate}
In the case of the Hyades this method gives
a distance of  37.6~$pc$ with an accuracy of $16\%$ in respect
to the guess value or  $19\%$ in respect to  $46.3~pc$
of  \cite {Wallerstein2000}.

\section{Theoretical relationships}

\label{theoretical}
In order to confirm or not the physical basis of 
formulas~(\ref{bct}) and (\ref{bvt}) 
we performed a Taylor-series expansion 
to the second order of the exact equations as 
given by the 
the Planck distribution for the colors ,
see  Section~\ref{sec_colours} , and for the
bolometric correction , see  Section~\ref{sec_bc}.
A careful analysis on the  numerical
results  applied to the Sun is reported 
in Section~\ref{sec_sun}.

\subsection{Colors versus Temperature}

\label{sec_colours}
The brightness of the radiation from a blackbody  is 
\begin{equation}
 B_{\lambda}(T)
=
\left ( \frac{2 h c^2} {\lambda^5} \right )
\frac  {1 }
       {\exp (\frac {hc }{\lambda kT}) -1 }
\label{leggeplancklambda}
\quad ,
\end{equation}
where  $c$ is the light velocity   ,
       $k$ the Boltzmann  constant ,
       $T$ the equivalent brightness temperature and 
       $\lambda$ the considered wavelength , 
see~formula~(13) in \cite{Planck_1901},or formula~(275) 
in~\cite{Planck_1959}, or formula (1.52) in
\cite{rybicki} , or formula~(3.52) in~\cite{Kraus1986}.
 
The color-difference , ($C_1$ - $C_2$) ,  can be expressed as 
\begin{equation}
(C_1-C_2)=m_{\mathrm 1} - m_{\mathrm 2}  = K  - 2.5  \log_{10}
\frac
{\int S_{\mathrm 1} I_{\lambda} d\lambda}
{\int S_{\mathrm 2} I_{\lambda} d\lambda}
\quad,
\label{color}
\end {equation}
where  $S_{\lambda}$ is the sensitivity function in the region
specified by the index $\lambda$~,
$K$ is a constant 
  and  $I_{\lambda}$ is the energy flux
reaching the earth.
We now define a sensitivity function for a pseudo-monochromatic
system 
\begin {equation}
S_{\lambda} = \delta (\lambda -\lambda_i)
\quad  i=U,B,V,R,I
\quad,
\end {equation}
where $\delta$ denotes the Dirac delta function.
In this
pseudo-monochromatic color 
system the color-difference  is  
\begin{equation}
(C_1-C_2)= K  - 2.5  \log_{10}\frac{\lambda_2^5}{\lambda_1^5}
\frac  {(\exp (\frac {hc }{\lambda_2 kT}) -1) } {(\exp (\frac {hc }{\lambda_1 kT}) -1) }
\label{exact}
\quad .
\end{equation}

\begin{table}[h]
      \caption[]{Johnson system }
      \label{ubv}
         \begin{tabular}{|c|c|}
            \hline
           \hline           
symbol        & wavelength  (\AA) \\ 
            \hline
U            &  3600       \\
B            &  4400       \\
V            &  5500        \\
R            &  7100        \\
I            &  9700       \\
            \hline
            \hline 
         \end{tabular}
   \end{table}

The previous expression for the color can be expanded 
through a Taylor series about the point $T =\infty$ 
or making the change of variable $ x=\frac{1}{T}$ ,
about the point $x=0$ . When the expansion order is 2 
we have
\begin{eqnarray}
(C_1-C_2)_{app} =
-\frac{5}{2}\,\ln  \left( {\frac {{{\it \lambda_2}}^{4}}{{{\it \lambda_1}}^{4}}}
 \right)  \frac {1} {\ln  ( 10)} 
\nonumber \\
-\frac{5}{4}\,{\frac {hc
 \left( {\it \lambda_1}-{\it \lambda_2} \right) }{{\it \lambda_2}\,{\it 
\lambda_1}\,k\ln  \left( 10 \right) T}}-{\frac {5}{48}}\,{\frac {{h}^{2}
{c}^{2} \left( {{\it \lambda_1}}^{2}-{{\it \lambda_2}}^{2} \right) }{{{
\it \lambda_2}}^{2}{{\it \lambda_1}}^{2}{k}^{2}\ln  \left( 10 \right) {T}^
{2}}}
\quad  ,
\label{taylor}
\end{eqnarray}
where the index $app$ means  approximated.
We now continue inserting the value of the physical 
constants as given by  CODATA~\cite{CODATA2005} 
and wavelength of the color as given by Table 15.6 in \cite{cox}
 and visible in 
 Table~\ref{ubv}.
The wavelength of U,B and V are exactly the same
of the multicolor photometric system defined by~\cite{Johnson1966},
conversely R(7000~$\AA$) and I(9000~$\AA$) as given 
by~\cite{Johnson1966}  are 
slightly different from the values here used. 
We now continue  parameterizing the color as  
\begin{equation}
(C_1-C_2)_{app} = a +\frac{b}{T} +\frac{d}{T^2}
\quad  .
\label {abd}
\end{equation}

Another important step is the  calibration of the color 
on the maximum temperature $T_{cal}$ of the reference tables.
For example for MAIN SEQUENCE V at  $T_{cal}=42000$
, see Table~15.7 in \cite{cox} , $(B-V)=-0.3$ 
and therefore  a constant should be added to formula~(\ref{taylor}) 
in order to obtain such a value.
With these recipes we obtain  , for example 
\begin{eqnarray}
(B-V) =
- 0.4243+ \frac{3543}{T} + \frac {17480000}{T^2}
 \\
 MAIN~SEQUENCE,V~when~-0.33<(B-V)<1.64.  \nonumber          
\label{bvmain} 
\end{eqnarray} 
The basic parameters  $b$ and $d$  for the four
colors here considered are reported in Table~\ref{ubv}.
The parameter $a$  when WHITE DWARF ,
 MAIN~SEQUENCE,~V, GIANTS,~III and 
SUPERGIANTS,~I  are considered is reported in Table~\ref{coefa} ,
conversely  the Table~\ref{coefbd} reports the coefficient
$b$ and $d$ that are in common to the classes of stars here considered.
\begin{table}[h]
      \caption[]{Coefficient a }
      \label{coefa}
         \begin{tabular}{|c|c|c|c|c|}
            \hline
           \hline           
~             & (B-V)      & (U-B)    & (V-R)  & (R-I)   \\
WHITE~ DWARF       , $T_{cal}$[K]=25200          &- 0.1981   & - 1.234  & &     \\
MAIN~SEQUENCE,~V  ,  $T_{cal}$[K]=42000          & - 0.4243   & - 1.297   & - 0.233      &- 0.395    \\
GIANTS~III        ,  $T_{cal}$[K]=5050           & - 0.5271   & - 1.156    &- 0.4294   &- 0.4417         \\
SUPERGIANTS~I     ,  $T_{cal}$[K]=32000          & - 0.3978    & - 1.276   & - 0.2621  &- 0.420         \\
            \hline
            \hline 
         \end{tabular}
   \end{table}
The WHITE DWARF  calibration is  made on the values of 
$(U-B)$ and  $(B-V)$   for Sirius~B, 
see Wikipedia (http://www.wikipedia.org/).

\begin{table}[h]
      \caption[]{Coefficients b and d }
      \label{coefbd}
      \[
         \begin{tabular}{|c|c|c|c|c|}
            \hline
           \hline           
~             & (B-V)      & (U-B)    & (V-R)  & (R-I)   \\
b            & 3543     & 3936   & 3201       & 2944   \\
d            &17480000    & 23880000&12380000   &  8636000   \\
            \hline
            \hline 
         \end{tabular}
      \]
   \end{table}

The Taylor expansion agrees  very well with the original function
and  Figure~\ref{f10} reports the difference between 
the exact function as given by the ratio of two exponential 
and the Taylor expansion in the $(B-V)$  case.
\begin{figure}
  \begin{center}
\includegraphics[width=10cm]{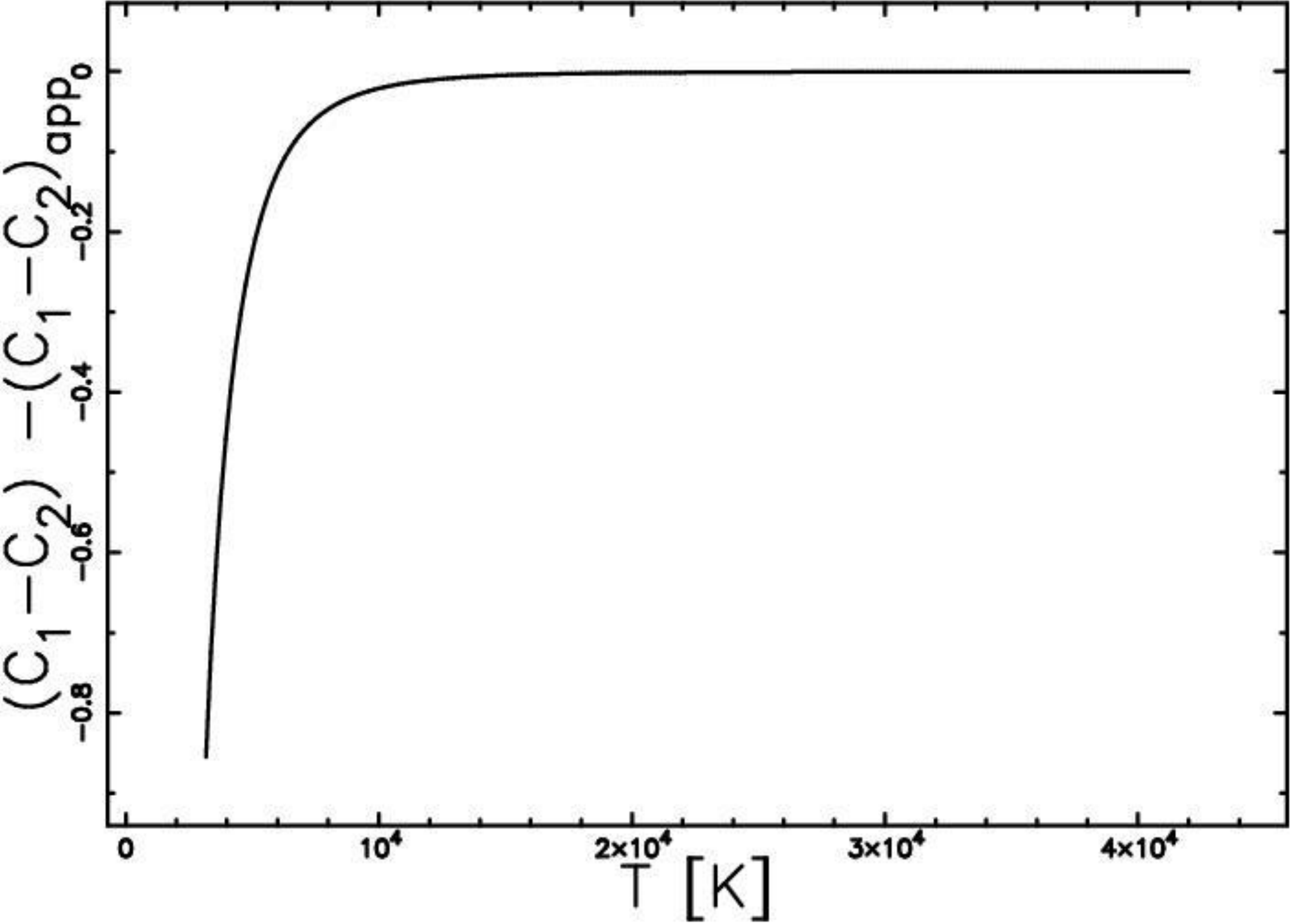}
  \end {center}
\caption {
Difference between $(B-V)$ , the exact value from the 
Planck distribution , 
and $(B-V)_{app}$ , approximate value as deduced from the
Taylor expansion for MAIN~SEQUENCE,~V 
          }%
    \label{f10}
    \end{figure}

In order to establish a range of reliability of the polynomial
expansion we solve the nonlinear equation
\begin{equation}
(C_1-C_2)  - (C_1-C_2)_{app} = f(T) = -0.4
\quad ,
\label{nonlinear}
\end{equation}
for $T$.
The solutions of the nonlinear equation
  are 
reported in Table~\ref{range} for the four colors here considered.
For the critical difference we have chosen the value $-0.4$ that approximately 
corresponds to $1/10$ of range of existence in  $(B-V)$ .
\begin{table}
      \caption[]{Range of existence of the Taylor expansion for 
                 MAIN~SEQUENCE,~V}
      \label{range}
      \[
         \begin{tabular}{|c|c|c|c|c|}
            \hline
           \hline           
            \noalign{\smallskip}
             & (B-V)    & (U-B)  & (V-R)  & (R-I)   \\
$T_{min}$ [K]   & 4137     &  4927  & 3413       &  2755   \\
$T_{max}$ [K]   & 42000    & 42000  & 42000      &  42000   \\
            \noalign{\smallskip}
            \hline
            \hline 
         \end{tabular}
      \]
   \end{table}
Figure~\ref{f11} reports the exact and the approximate
 value 
of $(B-V)$ as well as the calibrated data.

\begin{figure}
  \begin{center}
\includegraphics[width=10cm]{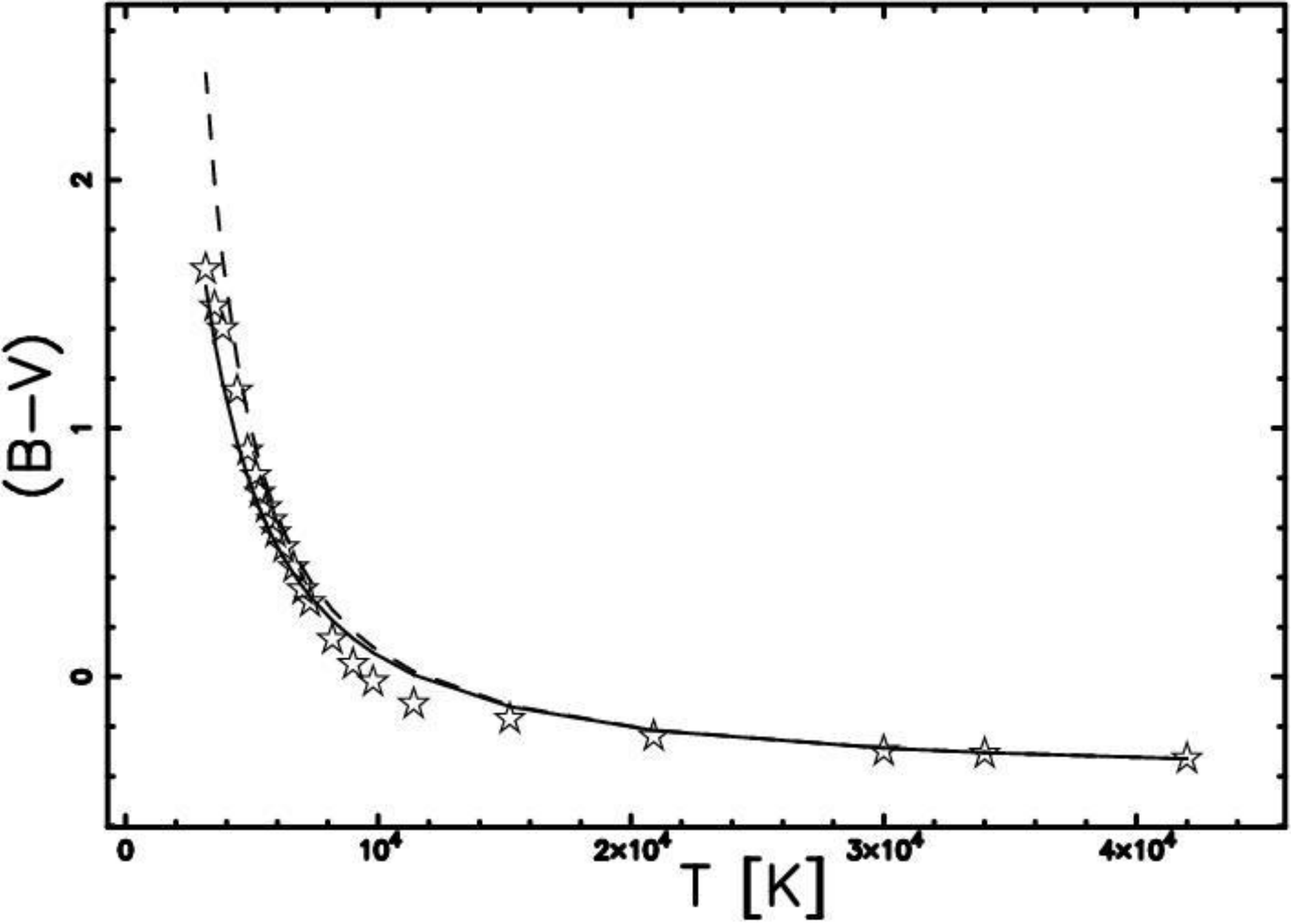}
  \end {center}
\caption {
Exact $(B-V)$  as  deduced from the Planck distribution , or
equation~(\ref{exact}),  traced with a full line.
Approximate  $(B-V)$ as deduced from the Taylor expansion, or 
equation~(\ref{taylor}),  traced with a dashed line.
The calibrated data for MAIN SEQUENCE V are extracted
from    Table 15.7 in~Cox~(2000) 
and are  represented through empty stars.
}%
    \label{f11}
    \end{figure}

\subsection{Bolometric Correction versus Temperature}

\label{sec_bc}
The  bolometric 
correction $BC$ , defined as always negative, is 
\begin {equation}
BC  = M_{\mathrm{bol}}  -M_{\mathrm{V}} 
\quad, \label{bc}
\end {equation}
where $M_{\mathrm{bol}}$   is the absolute bolometric magnitude
      and 
      $M_{V}$     is the absolute visual     magnitude.
It can be expressed as  
\begin{equation} 
BC=
\frac{5}{2} \frac{ ln \left (15 (\frac{hc}{kT\pi})^4  (\frac{1}{\lambda_V})^5 
\frac{1} {\exp(\frac{hc}{kT\lambda_V}) -1 }  
 \right )} 
{ln(10)}  +K_{BC}
\quad , 
\label{exactbc}
\end{equation}
where $\lambda_V$  is the  visual wavelength
and  $K_{BC}$ a constant.
We now  expand  
with  a Taylor series about the point $T =\infty$ 
\begin{eqnarray}
BC_{app}=   \nonumber\\
-\frac{15}{2}\,{\frac {\ln 
 \left( T \right) }{\ln  \left( 10 \right) }}-
\frac{5}{4}\,{\frac {hc}{k{\it 
\lambda_V}\,\ln  \left( 10 \right) T}}-{\frac {5}{48}}\,{\frac {{h}^{2}{
c}^{2}}{{k}^{2}{{\it \lambda_V}}^{2}\ln  \left( 10 \right) {T}^{2}}} +K_{BC}
\quad .
\end{eqnarray}
The constant $K_{BC}$ can be found   with the following 
procedure.
The maximum of  $BC_{app}$ is at $T_{max}$ , where
the index $max$ stands for maximum
\begin{equation}
T_{max}= 
\frac{1}{6}\,{\frac { \left( \frac{\sqrt {5}}{2}\,+\frac{1}{2}\right) ch}{k{\it \lambda_V}}}
\quad .
\end{equation}
Given the fact that the observed maximum in the $BC$ is -0.09 
at 7300~$K$  in the case of MAIN SEQUENCE V
we easily compute $K_{BC}$ and the following approximate
result is obtained
\begin{equation}
BC_{app}=
31.41- 3.257\,\ln  \left( T \right) - \frac{14200} {T} -
\frac{3.096~10^7}{T^2}
\label{appbc}
\quad .
\end{equation}
Figure~\ref{f12} reports the exact and the approximate
 value 
of $BC$ as well as the calibrated data.

\begin{figure}
  \begin{center}
\includegraphics[width=10cm]{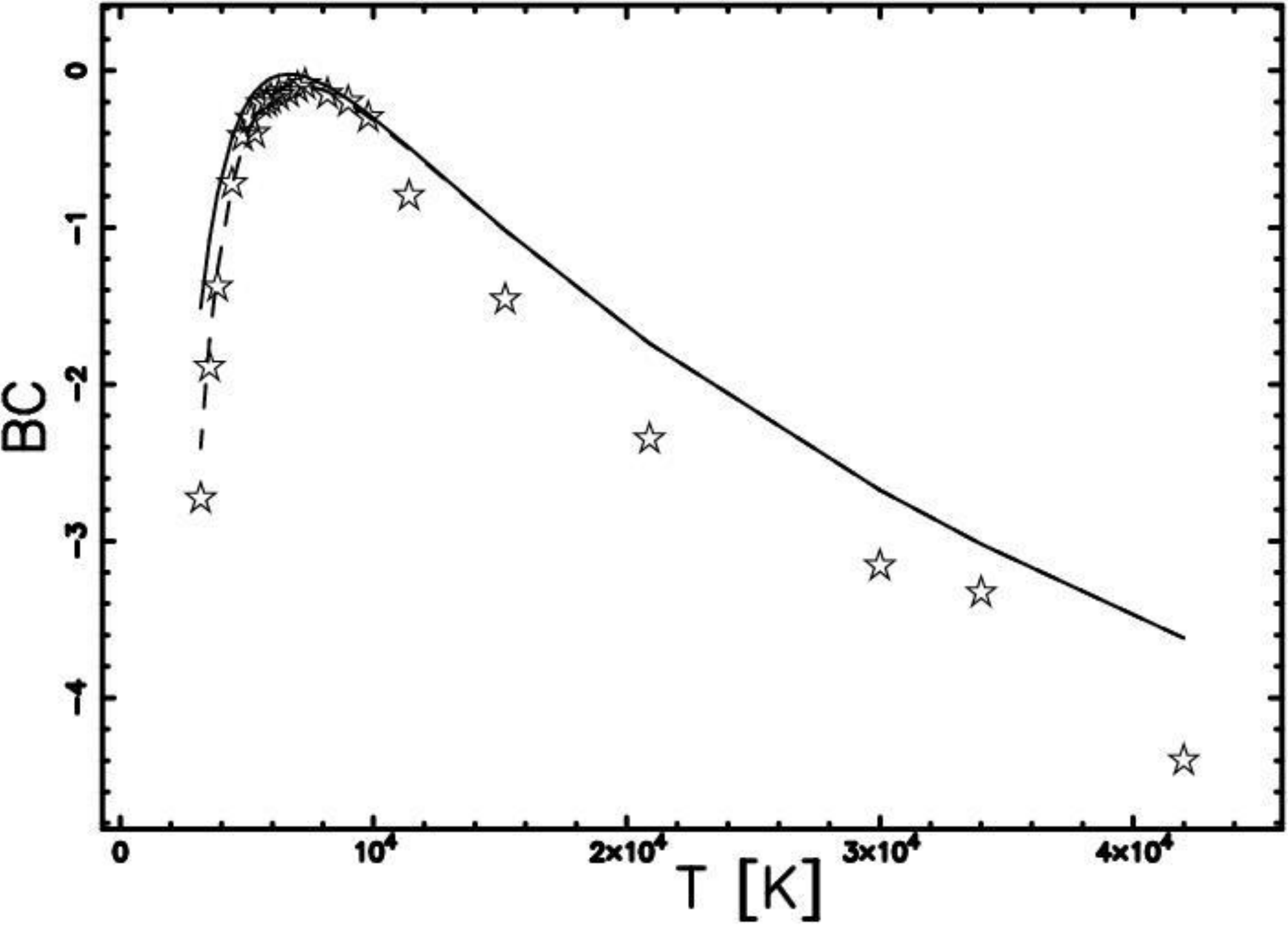}
  \end {center}
\caption {
Exact $BC$  as  deduced from the Planck distribution , or
equation~(\ref{exactbc}),  traced with a full line.
Approximate  $BC$ as deduced from the Taylor expansion, or 
equation~(\ref{appbc}),  traced with a dashed line.
The calibrated data for MAIN SEQUENCE V are extracted
from    Table 15.7 in~Cox~(2000) 
and are  represented through empty stars.
}%
    \label{f12}
    \end{figure}

The Taylor expansion agrees  very well with the original function
and  Figure~\ref{f13} reports the difference between 
exact function as given by  equation~(\ref{exactbc}) 
and the Taylor expansion  as given by equation~(\ref{appbc}).
\begin{figure}
  \begin{center}
\includegraphics[width=10cm]{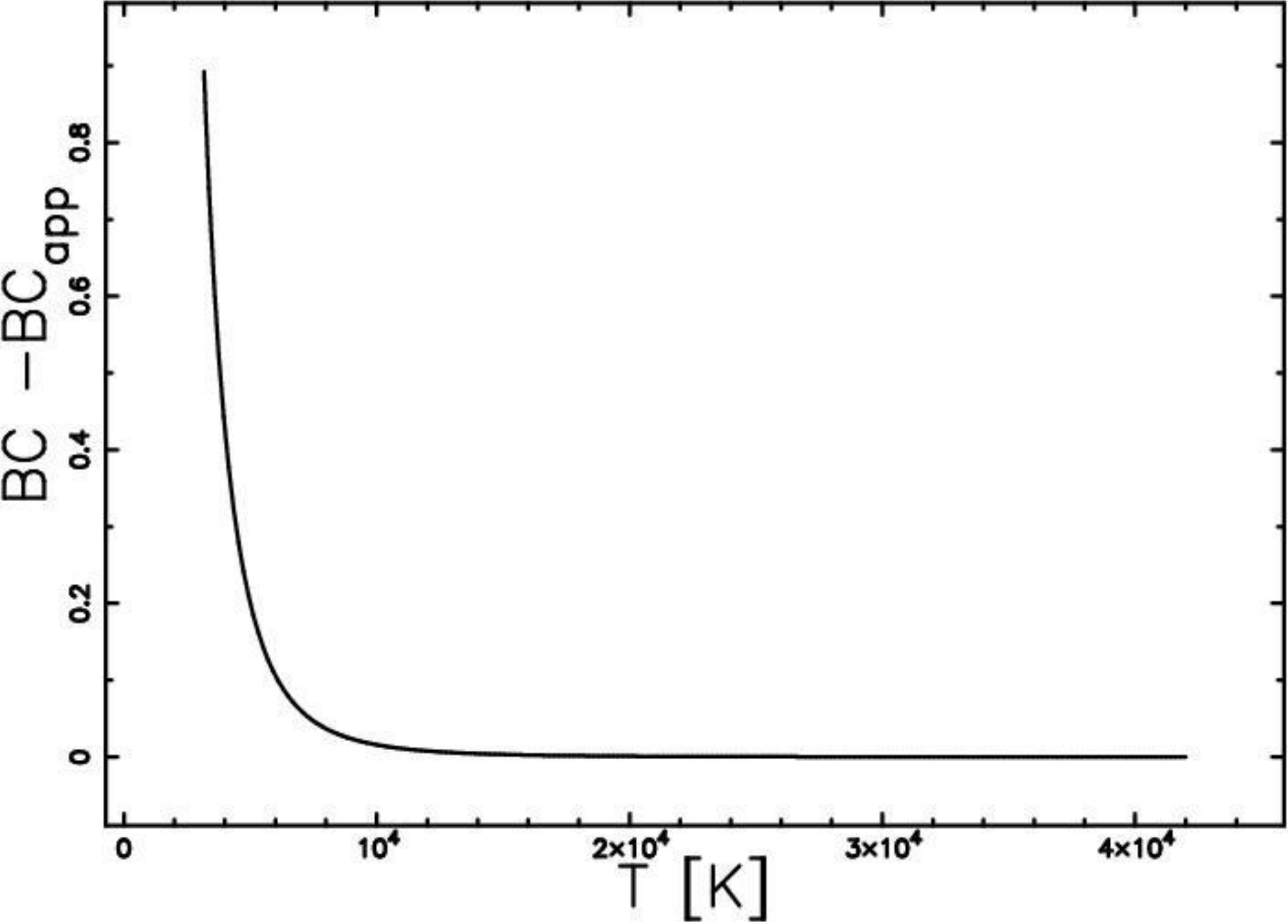}
  \end {center}
\caption {
Difference between $(B-V)$ , the exact value from the 
Planck distribution , 
and $(B-V)_{app}$ , approximate value as deduced from the
Taylor expansion for MAIN~SEQUENCE,~V 
          }%
    \label{f13}
    \end{figure}

In order to establish a range of reliability of the polynomial
expansion we solve the nonlinear equation
for $T$
\begin{equation}
BC  - BC_{app} =  -0.4
\quad .
\label{bcnonlinear}
\end{equation}
The solution of the previous nonlinear equation
allows to state that  the bolometric correction 
 as derived from a  Taylor 
expansion  for MAIN~SEQUENCE,~V 
is reliable in the range  $4074~K <T< 42000~K$.

\subsection{The Sun as a blackbody radiator }
\label{sec_sun}

The framework previously derived allows to compare our formulas
with one specific star of spectral type G2V with T=5777~K named
Sun. In order to make such a comparison we reported in Table~\ref{bvsun}
 the
various value of $(B-V)$ as reported by different methods as well as
in Table~\ref{risun}  the value of the infrared color (R-I). 
From a careful
examination of the two tables we conclude that our model works more 
properly in the  far-infrared window  in respect to the optical
one.

\begin{table}
      \caption[]{(B-V) of the Sun , T=5777~K}
      \label{bvsun}
      \[
         \begin{tabular}{|l|c|}
            \hline
           \hline           
            \noalign{\smallskip}
meaning           & (B-V)    \\ 
            \noalign{\smallskip}
            \hline
            \noalign{\smallskip}
calibration, ~Cox ~(2000)           &  0.65           \\
            \hline
            \noalign{\smallskip}
here, ~Taylor ~expansion             &  0.711     \\
here ,~Planck~ formula              &  0.57       \\
least~square~method, Zaninetti~(2005)      &  0.633      \\
Allen~ (1973)                       &  0.66        \\
Sekiguchi ~\&~ Fukugita~ (2000)     &  0.627        \\
Johnson ~ (1966)                    &  0.63          \\
            \noalign{\smallskip}
            \hline
            \hline 
         \end{tabular}
      \]
   \end{table}

\begin{table}
      \caption[]{(R-I) of the Sun , T=5777~K}
      \label{risun}
      \[
         \begin{tabular}{|l|c|}
            \hline
           \hline           
            \noalign{\smallskip}
meaning           & (R-I)    \\ 
            \noalign{\smallskip}
            \hline
            \noalign{\smallskip}
calibration, ~Cox ~(2000)            &  0.34          \\
            \hline
            \noalign{\smallskip}
here, ~Taylor ~expansion             &  0.37     \\
here ,~Planck~ formula               &  0.34       \\
            \noalign{\smallskip}
            \hline
            \hline 
         \end{tabular}
      \]
   \end{table}

\section{Conclusions}

{\bf New formulas~} A new analytical approach based on five basic equations
allows to connect the color $(B-V)$ of the stars
with the  absolute visual magnitude,  the  mass, the radius 
and the luminosity.The suggested method is based on eight parameters that can
be precisely derived from the calibration tables;
this is  the case of
MAIN  V  ,  GIANTS III and  SUPERGIANTS I.
In absence of calibration tables the eight parameters
can be derived mixing  four theoretical parameters  extracted
 fr\-om
Planck distribution  with four parameters that can be found
minimizing the  $\chi^2$ connected with the observed visual
magnitude ; this is the case of white dwarfs.
In the case of white dwarfs the
mass-luminosity relationship , see Table~\ref{coefficients_dwarf},
is
\begin{eqnarray}
\log_{10}(\frac {L}{L_{\sun}})  = 0.28 + 2.29 \log_{10}(\frac {{\mathcal M}}
 {{\mathcal M}_{\sun}})
\label{eqn_lm_white}
\quad,
\\
white~dwarf,   \quad   0.005 \mathcal M_{\sun} < \mathcal M  < 0.24 \mathcal M_{\sun}
\quad       .    \nonumber
\end{eqnarray}

{\bf Applications~} The applications of the new formulas to the open clusters
such as Hyades and Pleiades allows to speak of
universal laws for the star's main parameters.
In absence of accurate methods to deduce the distance of an open
cluster an approximate evaluation can be done.

{\bf Theoretical bases~} The reliability of an expansion at the second order of the 
colors and bolometric correction for stars
as derived from the 
Planck distribution
is carefully explored and the range of existence 
in temperature of the expansion is determined.

{\bf Inverse function~} In this paper we have chosen a simple 
hyperbolic behavior for  $(B-V)$ as function of the temperature
as  given by formula~(\ref{bvt}). This function can be easily 
inverted in order to obtain $T$ as function 
of $(B-V)$ (MAIN~SEQUENCE,~V)
\begin{eqnarray}
\label{simple}
T= 
\frac{7360}
{\it (B-V) + 0.641 } ~K
 \\
 MAIN~SEQUENCE,V~when~4137 < T[K] < 42000    \nonumber          
\\
or~when~-0.33<(B-V)<1.45  \nonumber          
\quad .
\end{eqnarray}

When conversely a more complex behavior is chosen ,
for example a two degree polynomial expansion  
in $\frac{1}{T}$ as given by formula~(\ref{taylor}) ,
the inverse formula that gives $T$ as function of
$(B-V)$ (MAIN~SEQUENCE,~V)   is more 
complicated than formula~(\ref{simple}),
\begin{eqnarray}
T  =   \nonumber \\
 5000\,{\frac { 0.355~10^8+\sqrt {{ 4.217\times 10^{15}}+{ 6.966
\times 10^{15}}\,{\it (B-V)}}}{ 10^8\,{\it (B-V)}+ 0.4244\, 10^8}}
\quad K
 \\
 MAIN~SEQUENCE,V~when~4137 < T[K] < 42000    \nonumber          
\\
or~when~-0.33<(B-V)<1.45  \nonumber          
\quad .
\end{eqnarray}
The mathematical treatment that allows to deduce the 
coefficients of the series reversion can be found 
in \cite{Morse_1953,Dwight_1961,Abramowitz1965}. 

\begin{thebibliography}{39}
\expandafter\ifx\csname natexlab\endcsname\relax\def\natexlab#1{#1}\fi

\bibitem[{{Abramowitz} \& {Stegun}(1965)}]{Abramowitz1965}
{Abramowitz}, M. \& {Stegun}, I.~A. {: 1965}, {Handbook of mathematical
  functions with formulas, graphs, and mathematical tables} (New York: Dover)

\bibitem[{{Al-Wardat}(2007)}]{Al-Wardat2007}
{Al-Wardat}, M.~A. {: 2007}, Astronomische Nachrichten, {\bf 328}, 63

\bibitem[{{Allen}(1973)}]{Allen1973}
{Allen}, C.~W. {: 1973}, {Astrophysical quantities} (London: University of
  London, Athlone Press, | 3rd ed.)

\bibitem[{{Bedding} {et~al.}(1998){Bedding}, {Booth}, \& {Davis}}]{Bedding1998}
{Bedding}, T.~R., {Booth}, A.~J., \& {Davis}, J., eds. 1998, {Proceedings of
  IAU Symposium 189 on Fundamental Stellar Properties: The Interaction between
  Observation and Theory}

\bibitem[{{Bouy} {et~al.}(2006){Bouy}, {Moraux}, {Bouvier}, {Brandner},
  {Mart{\'{\i}}n}, {Allard}, {Baraffe}, \& {Fern{\'a}ndez}}]{Bouy2006}
{Bouy}, H., {Moraux}, E., {Bouvier}, J., {et~al.} {: 2006}, \apj, {\bf 637},
  1056

\bibitem[{{Bowers} \& {Deeming}(1984)}]{deeming}
{Bowers}, R.~L. \& {Deeming}, T. {: 1984}, {Astrophysics. I and II} (Boston:
  {Jones and Bartlett })

\bibitem[{{Chandrasekhar}(1967)}]{Chandrasekhar_1967}
{Chandrasekhar}, S. {: 1967}, {An introduction to the study of stellar
  structure} (New York: Dover, 1967)

\bibitem[{{Chiosi} {et~al.}(1992){Chiosi}, {Bertelli}, \&
  {Bressan}}]{Chiosi1992}
{Chiosi}, C., {Bertelli}, G., \& {Bressan}, A. {: 1992}, \araa, {\bf 30}, 235

\bibitem[{{Cox}(2000)}]{cox}
{Cox}, A.~N. {: 2000}, {Allen's astrophysical quantities} (New York: Springer)

\bibitem[{{Dahn} {et~al.}(1982){Dahn}, {Harrington}, {Riepe}, {Christy},
  {Guetter}, {Kallarakal}, {Miranian}, {Walker}, {Vrba}, {Hewitt}, {Durham}, \&
  {Ables}}]{Dahn1982}
{Dahn}, C.~C., {Harrington}, R.~S., {Riepe}, B.~Y., {et~al.} {: 1982}, \aj,
  {\bf 87}, 419

\bibitem[{{de Bruijne} {et~al.}(2001){de Bruijne}, {Hoogerwerf}, \& {de
  Zeeuw}}]{Hyades2001}
{de Bruijne}, J.~H.~J., {Hoogerwerf}, R., \& {de Zeeuw}, P.~T. {: 2001}, \aap,
  {\bf 367}, 111

\bibitem[{{Dwight}(1961)}]{Dwight_1961}
{Dwight}, H.~B. {: 1961}, {Mathematical tables of elementary and some higher
  mathematical functions} (New York: Dover)

\bibitem[{{ESA}(1997)}]{Esa}
{ESA}. {: 1997}, VizieR Online Data Catalog, {\bf 1239}, 0

\bibitem[{{Hertzsprung}(1905)}]{Hertzsprung_1905}
{Hertzsprung}, E. {: 1905}, Zeitschrift für Wissenschaftliche Photographie,
  {\bf 3}, 442

\bibitem[{{Hertzsprung}(1911)}]{Hertzsprung_1911}
{Hertzsprung}, E. {: 1911}, Publikationen des Astrophysikalischen
  Observatoriums zu Potsdam, {\bf 63}

\bibitem[{{Johnson}(1966)}]{Johnson1966}
{Johnson}, H.~L. {: 1966}, \araa, {\bf 4}, 193

\bibitem[{{Kraus}(1986)}]{Kraus1986}
{Kraus}, J.~D. {: 1986}, {Radio astronomy} (Powell, Ohio: Cygnus-Quasar Books,
  1986)

\bibitem[{{Lang}(1999)}]{lang}
{Lang}, K.~R. {: 1999}, {Astrophysical formulae. (Third Edition)} (New York:
  Springer)

\bibitem[{{Madore}(1985)}]{Madore1985}
{Madore}, B.~F., ed. 1985, {Cepheids: Theory and observations; Proceedings of
  the Colloquium, Toronto, Canada, May 29-June 1, 1984}

\bibitem[{{Maeder} \& {Renzini}(1984)}]{Maeder1984}
{Maeder}, A. \& {Renzini}, A., eds. 1984, {Observational tests of the stellar
  evolution theory; Proceedings of the Symposium, Geneva, Switzerland,
  September 12-16, 1983}

\bibitem[{{McCook} \& {Sion}(1999)}]{McCook1999}
{McCook}, G.~P. \& {Sion}, E.~M. {: 1999}, \apjs, {\bf 121}, 1

\bibitem[{{Mermilliod} {et~al.}(1997){Mermilliod}, {Turon}, {Robichon},
  {Arenou}, \& {Lebreton}}]{Mermilliod1997}
{Mermilliod}, J.-C., {Turon}, C., {Robichon}, N., {Arenou}, F., \& {Lebreton},
  Y. 1997, in ESA SP-402: Hipparcos - Venice '97, 643--650

\bibitem[{{Micela} {et~al.}(1999){Micela}, {Sciortino}, {Harnden}, {Kashyap},
  {Rosner}, {Prosser}, {Damiani}, {Stauffer}, \& {Caillault}}]{Micela1999}
{Micela}, G., {Sciortino}, S., {Harnden}, Jr., F.~R., {et~al.} {: 1999}, \aap,
  {\bf 341}, 751

\bibitem[{{Mohr} \& {Taylor}(2005)}]{CODATA2005}
{Mohr}, P.~J. \& {Taylor}, B.~N. {: 2005}, Reviews of Modern Physics, {\bf 77},
  1

\bibitem[{{Morgan} \& {Keenan}(1973)}]{MK1973}
{Morgan}, W.~W. \& {Keenan}, P.~C. {: 1973}, \araa, {\bf 11}, 29

\bibitem[{{Morse} \& {Feshbach}(1953)}]{Morse_1953}
{Morse}, P.~H. \& {Feshbach}, H. {: 1953}, Methods of Theoretical Physics (New
  York: Mc Graw-Hill Book Company)

\bibitem[{{Padmanabhan}({2001})}]{Padmanabhan_II_2001}
{Padmanabhan}, P. {: {2001}}, {Theoretical astrophysics. Vol. II: Stars and
  Stellar Systems} ({Cambridge, MA}: {Cambridge University Press})

\bibitem[{{Planck }(1901)}]{Planck_1901}
{Planck }, M. {: 1901}, Annalen der Physik, {\bf 309}, 553

\bibitem[{{Planck }(1959)}]{Planck_1959}
{Planck }, M. {: 1959}, {The theory of heat radiation} (New York: Dover
  Publications)

\bibitem[{{Renzini} \& {Fusi Pecci}(1988)}]{Renzini1988}
{Renzini}, A. \& {Fusi Pecci}, F. {: 1988}, \araa, {\bf 26}, 199

\bibitem[{{Rosenberg}(1911)}]{Rosenberg_1911}
{Rosenberg}, H. {: 1911}, Astronomische Nachrichten, {\bf 186}, 71

\bibitem[{{Russell}(1914{\natexlab{a}})}]{Russell1914Nature}
{Russell}, H.~N. {: 1914{\natexlab{a}}}, \nat, {\bf 93}, 252

\bibitem[{{Russell}(1914{\natexlab{b}})}]{Russell_1914_a}
{Russell}, H.~N. {: 1914{\natexlab{b}}}, The Observatory, {\bf 37}, 165

\bibitem[{{Russell}(1914{\natexlab{c}})}]{Russell_1914_b}
{Russell}, H.~N. {: 1914{\natexlab{c}}}, Popular Astronomy, {\bf 22}, 331

\bibitem[{{Rybicki} \& { Lightman}(1985)}]{rybicki}
{Rybicki}, G. \& { Lightman}, A. {: 1985}, Radiative Processes in Astrophysics
  (New-York: Wiley-Interscience)

\bibitem[{{Stern} {et~al.}(1995){Stern}, {Schmitt}, \& {Kahabka}}]{Stern1995}
{Stern}, R.~A., {Schmitt}, J.~H.~M.~M., \& {Kahabka}, P.~T. {: 1995}, \apj,
  {\bf 448}, 683

\bibitem[{{Vogt}(1926)}]{Vogt_1926}
{Vogt}, H. {: 1926}, Astronomische Nachrichten, {\bf 226}, 301

\bibitem[{{Wallerstein}(2000)}]{Wallerstein2000}
{Wallerstein}, G. 2000, in Bulletin of the American Astronomical Society,
  102--+

\bibitem[{{Zaninetti}(2005)}]{zaninetti05}
{Zaninetti}, L. {: 2005}, Astronomische Nachrichten, {\bf 326}, 754

\end{thebibliography}

\end{document}